# From Brussels Effect to Gravity Assists: Understanding the Evolution of the GDPR-Inspired Personal Information Protection Law in China

Wenlong Li[a]
Jiahong Chen[b]
[a] Edinburgh Centre for Data, Culture & Society, United Kingdom. wenlong.li@ed.ac.uk
[b] School of Law, University of Sheffield, United Kingdom. jiahong.chen@sheffield.ac.uk

**Abstract**

This paper explores the evolution of China's Personal Information Protection Law (PIPL) and situates it within the context of global data protection development. It draws inspiration from the theory of 'Brussels Effect' and provides a critical account of its application in non-Western jurisdictions, taking China as a prime example. Our objective is not to provide a comparative commentary on China's legal development but to illuminate the intricate dynamics between the Chinese law and the EU's GDPR. We argue that the trajectory of China's Personal Information Protection Law calls into question the applicability of the Brussels Effect: while the GDPR's imprint on the PIPL is evident, a deeper analysis unveils China's nuanced, non-linear adoption that diverges from many assumptions of the Brussels Effect and similar theories. The evolution of the GDPR-inspired PIPL is not as a straightforward outcome of the Brussels Effect but as a nuanced, intricate interplay of external influence and domestic dynamics. We introduce a complementary theory of 'gravity assist', which portrays China's strategic instrumentalisation of the GDPR as a template to shape its unique data protection landscape. Our theoretical framework highlights how China navigates through a patchwork of internal considerations, international standards, and strategic choices, ultimately sculpting a data protection regime that has a similar appearance to the GDPR but aligns with its distinct political, cultural and legal landscape. With a detailed historical and policy analysis of the PIPL, coupled with reasonable speculations on its future avenues, our analysis presents a pragmatic, culturally congruent approach to legal development in China. It signals a trajectory that, while potentially converging at a principled level, is likely to diverge significantly in practice, driven by China's broader socio-political and economic agendas rather than foundational premises of EU data protection law and its global aspirations. It thus indicates the inherent limitations of applying Brussels Effect and other theoretical frameworks to non-Western jurisdictions such as China, while highlighting the imperative for integrating complementary theories to more accurately navigate complex legal landscapes.

# 1. Introduction

Data protection laws have risen to global prominence as vital instruments for navigating the complexities of the information age while shaping international policy landscapes.[1] At the forefront of this landscape stands the EU's General Data Protection Regulation (GDPR),[2] an epochal piece of legislation of the EU with far-reaching influence beyond borders, jurisdictions and cultures, propelling data protection to the centre of global discourse.[3] The resonance of the GDPR reverberated far beyond the EU's borders, initiating discussions on the EU's externalisation of laws. Along a decades-old line of theoretical scholarship,[4] the Brussels Effect, advanced by Anu Bradford, stands prominently, which posits that the EU's regulations possess the uncanny ability to influence and shape regulatory standards worldwide, generating a global regulatory pull towards the EU standards.[5] It has gained global recognition as a framework to comprehend the development of data protection worldwide, particularly in the context of the EU's de facto externalisation of its laws, attributed mainly to the EU's regulatory capacity and ideation powers.[6]

Yet, the Brussels Effect along the lines of theories on international data protection development has encountered intriguing challenges when applied to non-Western contexts.[7] While it rightly highlights

---

[1] Anupam Chander and Paul M Schwartz, 'Privacy and/or Trade' (2023) 90(1) University of Chicago Law Review 49. See also Graham Greenleaf, 'Global data privacy laws 2023: 162 national laws and 20 Bills' (2023) <https://dx.doi.org/10.2139/ssrn.4426146> accessed 20 October 2023; Jackson Adams and Hala Almahmoud, 'The Meaning of Privacy in the Digital Era' (2023) 15(1) International Journal of Security and Privacy in Pervasive Computing (IJSPPC) 1. In this paper, we use the concepts of 'data protection', 'data privacy', and 'personal information protection' interchangeably with recognition of their nuanced differences. The concept of data protection is used mainly in contexts regarding the EU and its member states, whereas data privacy is used in general contexts. As China deliberately selects the concept of 'personal information protection', we use this concept when the Chinese law is concerned.

[2] Regulation (EU) 2016/679 of the European Parliament and of the Council of 27 April 2016 on the protection of natural persons with regard to the processing of personal data and on the free movement of such data, and repealing Directive 95/46/EC (General Data Protection Regulation) [2016] OJ L119/1 ('GDPR').

[3] Graham Greenleaf, 'Global Data Privacy 2023: DPA Networks Almost Everywhere' (2023) <https://dx.doi.org/10.2139/ssrn.4461729> accessed 20 October 2023; Graham Greenleaf, 'Global data privacy laws: EU leads US and the rest of the world in enforcement by penalties' (2023) <https://dx.doi.org/10.2139/ssrn.4409491> accessed 20 October 2023; Graham Greenleaf, 'Global data privacy laws 2023: International standards stall, but UK disrupts' (2023) <https://dx.doi.org/10.2139/ssrn.4530145> accessed 20 October 2023.

[4] See, for example, Colin J. Bennett and Charles D. Raab, *The Governance of Privacy: Policy Instruments in Global Perspective* (The MIT Press 2006); Graham Greenleaf, 'The influence of European data privacy standards outside Europe: implications for globalization of Convention 108' (2012) 2(2) International Data Privacy Law 68; Lee A Bygrave, 'Privacy and data protection in an international perspective' (2010) 56(8) Scandinavian studies in law 165.

[5] Anu Bradford, 'The Brussels Effect' (2012) 107(1) Northwestern University Law Review 1; Anu Bradford, *The Brussels Effect: How the European Union Rules the World* (Oxford University Press 2020).

[6] Bradford, 'The Brussels Effect' (n 5); Bradford, *The Brussels Effect* (n 5). Cf. see Lee A Bygrave, 'The "Strasbourg Effect" on data protection in light of the "Brussels Effect": Logic, mechanics and prospects' (2021) 40 Computer Law & Security Review 105460. More recently, the discussion on the Brussels Effect has extended to new digital regulatory areas such as AI governance. See Charlotte Siegmann and Markus Anderljung, 'The Brussels effect and artificial intelligence: How EU regulation will impact the global AI market' (2022) <https://doi.org/10.48550/arXiv.2208.12645> accessed 20 October 2023; Graham Greenleaf, 'The 'Brussels effect'of the EU's 'AI Act'on data privacy outside Europe' (2021) <https://ssrn.com/abstract=3898904> accessed 20 October 2023.

[7] We use the concept of 'Western', with due cognition of its conceptual ambiguity and political implications, to refer to developed countries in the Global North that first participated in the development of national and international data privacy frameworks. These include, notably, the United States, the majority of European countries, Australia, New Zealand, and some Asian countries that are largely 'Westernised' e.g. Singapore. It is acknowledged, however, that critical differences exist between these regimes about the level and means of data privacy, and that conflicting propositions do exist even in one country (e.g. the state-level and federal regimes in

the EU's obvious influence on global regulations, this influence may not be equally pronounced across all countries and regions, as Bradford and other scholars explicitly recognise. For instance, after analysing the external and internal dynamics around data privacy policymaking in a country–as we aim to in this paper with the case of China–one might come to the conclusion that the European model's impact on regions with closer economic and political ties to the EU might manifest itself differently from how it does on developing countries, particularly with distinct cultural, legal and economic contexts. Setting aside the legitimacy questions about this regulatory influence and power over other jurisdictions,[8] the applicability of the prevailing theories might be impacted by such factors as trade relationships, local norms and values, regulatory capacity, legal traditions, etc.[9] Regions such as East Asia and parts of the Global South grapple with the intricate interplay of local norms, geopolitical dynamics, and global standards, while struggling with a sub-optimal level of democracy, rule of law, and fundamental rights that the EU holds dear.[10]

The emergence of China's Personal Information Protection Law (PIPL) [11] has punctuated and complicated this conversation, defying simplistic categorisation and characterisation in relation to the Brussels Effect framework. China's data privacy journey appears to have embraced the EU's regulatory approach, sparking considerable policy and theoretical interest. Yet, we argue that it is actually a paradoxical blend of alignment and distinction, intricately and slowly woven into its cultural, political and legal fabric. While echoing the concepts and principles of the GDPR, the PIPL presents stark departures and even challenges the conventional premises of regulatory convergence. This contrast between alignment and distinction has been reflected in numerous detailed textual comparisons and other legal analyses. [12] Yet, the intrinsic complexities of data protection evolution are largely unexamined within distinct geopolitical and cross-cultural contexts.

The PIPL, we acknowledge, has already been subject to moderate examination through legal and theoretical lenses. For instance, Jia makes a theoretical case for China's unconventional move to proactively legislate for privacy despite its political, cultural and ideological aversion.[13] Creemers conceptualises China's model to be built upon two pillars—personal information protection and data security—considering the latter to be 'a considerable innovation'.[14] You points out critical deficits of China's personal information protection law but nonetheless contends that 'half a loaf is better than

---

the US). On the other hand, 'non-Western' is used in this paper to refer to those countries with apparent political and/or cultural differences, which happen to be developing countries in Asia, Africa and the Global South, characteristically the second-movers in the realm of data privacy laws.

[8] Cara Mannion, 'Data imperialism: The GDPR's disastrous impact on Africa's E-commerce markets' (2020) 53 Vanderbilt Journal of Transnational Law 685.

[9] Bradford, 'The Brussels Effect' (n 5).

[10] See Tom Ginsburg, 'Does Law Matter for Economic Development? Evidence from East Asia' (2000) 34 Law and Society Review 829; Randall Peerenboom, 'Globalization, path dependency and the limits of law: Administrative law reform and rule of law in the People's Republic of China' (2001) 19 Berkeley Journal of International Law 161; Kanishka Jayasuriya, 'The rule of law and governance in East Asia' in Mark Beeson (ed) *Reconfiguring East Asia* (Routledge 2002) 99-116; Shucheng Wang, 'Authoritarian Legality and Legal Instrumentalism in China' (2022) 10(1) The Chinese Journal of Comparative Law 154.

[11] Personal Information Protection Law (China) (中华人民共和国个人信息保护法) [2021] ('PIPL'). Unless otherwise stated, all translation from Chinese was performed by the authors. The translation into English of direct quotes and titles of primary and secondary sources is accompanied by the original Chinese texts wherever necessary.

[12] See, for example, Julia Zhu, 'The Personal Information Protection Law: China's Version of the GDPR' (2022) <https://www.jtl.columbia.edu/bulletin-blog/the-personal-information-protection-law-chinas-version-of-the-gdpr> accessed 20 October 2023; Xu Ke and others, 'Analyzing China's PIPL and how it compares to the EU's GDPR' (2021) <https://iapp.org/news/a/analyzing-chinas-pipl-and-how-it-compares-to-the-eus-gdpr> accessed 20 October 2023.

[13] Mark Jia, 'Authoritarian Privacy' (2023) <https://dx.doi.org/10.2139/ssrn.4362527> accessed 20 October 2023.

[14] Rogier Creemers, 'China's emerging data protection framework' (2022) 8(1) Journal of Cybersecurity 1.

none'.[15] More earlier characterisations of China's approach to data privacy, for instance by Pernot-Leplay,[16] Geller,[17] De Hert and Papakonstantinou[18], are largely dated as the PIPL comes into play. Overall, the existing scholarship has contributed by presenting the nuanced landscape of data privacy in China but has not, we argue, provided a coherent, concise, and theory-grounded analysis of the new Chinese law. More importantly, little has been done to situate legal development in China within the existing conceptual and theoretical frameworks to aid our understanding of it, and particularly its potential European roots.

Against this backdrop, this paper intends to grapple with these paradoxes presented by China's legislative evolution while reflecting on the relevance and application of main theoretical frameworks, the Brussels Effect in particular. It particularly unpacks the intricate relationship between the GDPR and the PIPL, raising critical questions about the extent of alignment and the complex motivations driving China's personal information protection law. Our paper rests upon a critical reflection on the Brussels Effect as applied to non-Western contexts, extending its conceptual boundaries to accommodate and appreciate local exigencies. Following a critical analysis of the Brussels Effect and its limits, we introduce the theory of 'Gravity Assist', inspired by the orbital mechanics of space exploration, which not only accounts for the global influences but also delves into the interplay of China's internal dynamics. The theory is intended to develop a nuanced perspective, shedding light on the relevance of China's distinctive cultures and politics to the evolution of personal information protection law in addition to the EU influences. Gravity Assist unravels the complexities of this legal evolution, demonstrating how a country's unique context and priorities, often understated in global or Eurocentric frameworks, interact with external influences to co-shape its data protection landscape. The concept captures the deliberate and strategic processes of absorbing, adapting, and assimilating established data protection frameworks in China and other non-Western countries. Overall, Gravity Assist, we argue, represents a more advanced framework that explains China's measured approach to privacy legislation, its strategic experimentation with various legal domains, its nuanced convergence and divergence with the GDPR, and its future trajectories guided by the overarching cyber sovereignty agenda. The theory, we believe, has broader implications for other countries in the Global South, potentially contributing also to the discussion of global data governance.

The contribution of this paper is threefold. First, we critically scrutinise the Brussels Effect's applicability beyond its conventional boundaries *i.e.,* mostly Western and developed countries. Acknowledging the caveats and limitations of this theory in non-Western contexts, we paint a more contextualised picture of China's journey to data protection development, underscoring the need to appreciate the complexities of regions with distinctive cultural, legal and economic contours, in addition to those of global data governance. Second, we develop the theory of 'Gravity Assist' that extends the Brussels Effect's gaze and shed light on the multifaceted interplay between external influences and local exigencies in shaping data protection laws. It serves not merely as a theoretical construct but as a roadmap guiding us through the complex pathways of legislative journeys in China and possibly other non-Western countries. Third, our examination of the Sino–EU dynamic underscores the complexity of the EU regulatory influence upon non-Western jurisdictions, sparking reflections on the convergence of data privacy at regional and global levels.

---

[15] Chuanman You, 'Half a loaf is better than none: The new data protection regime for China's platform economy' (2022) 45 Computer Law & Security Review 105668.

[16] Emmanuel Pernot-LePlay, 'China's approach on data privacy law: A third way between the US and the EU?' (2020) 8(1) Penn State Journal of Law & International Affairs 49.

[17] Anja Geller, 'How Comprehensive Is Chinese Data Protection Law? A Systematisation of Chinese Data Protection Law from a European Perspective' (2020) 69(12) GRUR International 1191.

[18] Paul De Hert and Vagelis Papakonstantinou, *The data protection regime in China: In-depth analysis for the LIBE Committee* (2015) <https://www.europarl.europa.eu/RegData/etudes/IDAN/2015/536472/IPOL_IDA(2015)536472_EN.pdf> accessed 20 October 2023.

This paper proceeds with four parts. After introduction, Section 2 provides a critical account of the Brussels Effect, including its assumptions and its contested applicability in non-Western contexts, with a particular reference to China. Section 3 introduces the theory of Gravity Assist and explains the extent to which it complements and refines the Brussels Effect. Section 4 situates the theory of gravity assist within China's socio-legal contexts, contemplating the past failed attempts of legislation, the present instrumentalisation of the GDPR as a template, and potential divergences and departure in the near future. Section 5 concludes the article.

## 2. Brussels Effect: A Critical Application and Contestation in China's Context

### 2.1. Brussels Effect and Beyond: A Theoretical Overview

In the legal and regulatory scholarship, there has been an established line of work investigating how and why international influences exist between data privacy regimes. As early as 1992, Bennett examined the similarities and differences between privacy laws of several Western countries and identified five plausible explanations for global convergence: (a) technological determinism (due to common technical issues); (b) emulation (adoption of other countries' laws); (c) elite networking (through a cross-national policy community); (d) harmonisation (by international organisations); and (e) penetration (by coercing a different state to conform to its own laws).[19] Two decades later, Greenleaf adds a six explanation to the list, i.e., (f) economic force by multinational non-state actors (i.e., the 'big techs'), taking into account the socio-technical landscape in which data protection norms are rapidly developed as a response to the urgent problems arising from new technologies or business practices.[20] These explanations are helpful in reaching a general understanding of international data privacy frameworks but less effective in explaining why a particular regime is more influential than others, and how exactly one country picks up a particular model of regulation.

Yet, the extent to which the 'European model' is the leading driving force is not incontestable, especially when other major powers may also present their own approaches as competing regulatory models. This raises the question of where the convergence of global data protection laws—if it ever happens—will be heading towards. Bennett and Raab contrast the 'race-to-the-bottom' theory with the 'race-to-the-top' (or 'trading-up') theory, alternatively known in the American context as 'Delaware effect' vs 'California effect' (based on the fact that regulation in California is more stringent than that in Delaware).[21] In reality, the EU appears to drive the overall standards of data protection upwards, with its GDPR hailed as the most comprehensive and stringent data protection law in history, setting an example for an increasing number of countries to follow over the last 15 years.[22]

How exactly the EU exerts influence on other countries on data protection matters is a popular site of inquiry, on which Kuner lists four mechanisms of the 'Europeanisation of internet regulation': (a) emulation and learning; (b) international negotiation; (c) coercion and conditionality (e.g. making access conditional on rules unilaterally set); and (d) blocking recognition of third country legal measures (e.g. limiting the enforceability of third-country courts).[23] In part, Kuner's framework describes how the EU's global regulatory powers have been generated through bi- or multilateral interactions (e.g. (a) and (b)) but also unilateral actions pursued by Brussels (e.g. (c) and (d)). Kuner hints at why policymakers in other countries might wish to—or, as the case might be, would have to—align their own systems with the EU's, but the motivations of 'influenced' regimes are not the central focus of his theory.

---

[19] Colin J Bennett, *Regulating privacy: Data protection and public policy in Europe and the United States* (Cornell University Press 1992) viii.

[20] Graham Greenleaf, *Asian Data Privacy Laws: Trade and Human Rights Perspectives* (Oxford University Press 2014).

[21] Bennett and Raab (n 4) 136.

[22] Greenleaf, *Asian Data Privacy Laws: Trade and Human Rights Perspectives* (n 20) 559. See also Greenleaf, 'Global data privacy laws 2023: 162 national laws and 20 Bills' (n 1).

[23] Christopher Kuner, 'The Internet and the Global Reach of EU Law' in Marise Cremona and Joanne Scott (eds), *EU Law Beyond EU Borders: The Extraterritorial Reach of EU Law* (Oxford University Press 2019) 130-134.

The influence of the EU's legislation across the globe, with data protection as a noticeable example, has been examined in greater depth in Anu Bradford's work, known as the Brussels Effect.[24] Bradford distinguishes two forms of EU influence, one directly affecting business practices of foreign corporations without the law of their home country being changed ('*de facto* Brussels Effect') and the other affecting the laws of other countries ('*de jure* Brussels Effect'). In this framework, a set of mechanisms are identified for the EU's global regulatory influence over other countries, including unilateral regulatory globalisation, treaties and institutions, and European courts and foreign litigants.[25] Bradford identifies three rationales to explain why non-European jurisdictions would adopt GDPR-like data privacy frameworks, including (a) the economic incentive to establish data free flows with the EU; (b) the already *de facto* compliance by international corporations; and (c) ideational power of the GDPR's presence as the 'gold standard'.[26] When it comes to China, however, Bradford herself has questioned, prior to the enactment of the PIPL, the significance that the *de jure* Brussels Effect holds in that country.[27] In a more recent book after the Brussels Effect, Bradford further compares the three 'digital empires', with China characterised as exerting its global influence through digital infrastructure and less active in setting data privacy standards.[28] This naturally raises the question as to how much Brussels Effect and other theories outlined above can explain the convergence between the GDPR and the PIPL, even if that phenomenon is just an instance of the Brussels Effect on paper.

**2.2. Brussels Effect in China Contested**

The difficulties in applying Brussels Effect and other related theories to the PIPL can be seen more clearly when their arguments are presented and evaluated as a whole. In answering the question of why certain countries have aligned their data privacy laws at least partly with those of a different jurisdiction, the accounts offered by these theories can be loosely grouped into two main aspects.

2.2.1. Technological and normative aspects

Despite the differences in cultural and legal traditions, the technical challenges posed by new technologies tend to be more common (Bennett's point (a)). As a result, there is scope for the later entrants to learn and emulate the laws of the earlier ones in this area (Bennett's point (b) and Kuner's point (a)), especially when the target regime is considered normatively superior, such as the GDPR (Bradford's point (c)). In Bygrave's words, they are somehow subscribed to the EU's normative or ideational power.[29]

It is true that some of the technical issues are shared in both the EU and China, and that the political appeal of the GDPR lends itself well to the legitimacy of any efforts to regulate the use of personal information in China. The normative power of the GDPR, as Bygrave contends, could be leveraged to justify a new regime that is similar at least at face value.[30] Indeed, in pre-PIPL times, some experts in China have cited the EU's data protection legal framework to justify their legislative suggestions made to policymakers (although they also cited the American approach as a competing model).[31] The technical maturity of the GDPR may offer significant legislative convenience to drafters of the PIPL, but the motivation of Chinese policymakers to adopt the expert advice of 'learning' from the European model is not reflected in the theories mentioned above.

---

[24] Bradford, 'The Brussels Effect' (n 5); Anu Bradford, 'Exporting standards: The externalization of the EU's regulatory power via markets' (2015) 42 International Review of Law and Economics 158; Bradford, *The Brussels Effect* (n 5).
[25] Bradford, *The Brussels Effect* (n 5) 67-78.
[26] ibid 147-155.
[27] 'China's practice of deploying data as a tool for social control is, in practice, a stark reminder that any de jure Brussels Effect on paper does not necessarily mean that the EU's regulations and principles are deployed in any meaningful way.' See ibid 154.
[28] Anu Bradford, *Digital Empires: The Global Battle to Regulate Technology* (Oxford University Press 2023).
[29] Bygrave, '"Strasbourg Effect"' (n 6).

The legal emulation and ideational power arguments are especially weak in the case of China, where the differences between the two jurisdictions have been evident in the recent Sino–European geopolitical climate. China's insistence, or at least appearing to insist, on its 'sovereignty' in cyberspace (which we will further discuss below in Sections 4.1 and 4.3) makes it an even less plausible hypothesis that China 'looked up' to the European model as the 'gold standard' or 'best practice'. Further elaboration is needed to explain why, despite not being a political ally, China's policymakers actively chose to adopt *some* of the components (and not others) from the EU's framework.

2.2.2. Economic and institutional aspects

The theories discussed above also give credit to the economic incentives for international alignment of data protection, as the fragmentation of rules governing the use of data would create barriers to cross-border data flows and frustrate international trade in data-intensive sectors. At the core of Bradford's framework, for instance, lies the creation of a common market, an overarching objective that may be shared with third countries that benefit from it.[32] This rationale of economic integration is also seen in Bennett and Raab's claims, citing Vogel's research, that richer and more powerful countries like the EU are incentivised to set harmonised standards.[33] State actors have the shared goal to reduce the level of divergence, whether through international negotiation and treaties (Bennett's point (c) and Kuner's point (b)), harmonisation by international organisations (Bennett's point (d)), fulfilment of transfer conditions or securing an 'adequacy decision' (Kuner's point (c) and Bradford's point (a)), or influence by multinational corporations (Greenleaf's point (f) and Bradford's point (b)). These mechanisms are not always multi- or bilateral and can sometimes be unilaterally imposed by a state in an economically more powerful position (Bennett's point (e) and Kuner's point (d)).

Indeed, there can be third countries in a desire to obtain an adequacy decision from the EU, resulting from the free flow of personal data towards that country. This however may not hold true in the case of China. Despite a degree of some trade interest, the significance of economic considerations in data privacy policy making is largely questionable, at least compared to other economies. China appears to take a pragmatic approach to deal with frictions[34], and there does not seem to be a genuine interest on the part of China to strike a trade deal with the EU in the digital services sector. While the two economic giants reached a trade agreement in 2020, it focuses mainly on the manufacturing sector.[35] Unlike the US, China is not a major exporter of digital services to the EU,[36] which gives the country a much weaker incentive to align its data privacy law in an attempt to facilitate the free flow of personal data from the EU.

---

[30] ibid 6-8.

[31] See Hanhua Zhou, *Personal Information Protection Law (Expert Proposal) and Legislative Research Report (个人信息保护法（专家建议稿）及立法研究报告)* (Law Press China 2006); Xinbao Zhang, 'A Discussion on the Legislation of Individual Information Protection of China (我国个人信息保护法立法主要矛盾研讨)' (2018) 58(5) Jilin University Journal Social Sciences Edition 45; Aimin Qi, 'Scholarly Proposal of the Personal Information Protection Law (中华人民共和国个人信息保护法学者建议稿)' (2019) 37(1) Hebei Law Science 33.

[32] Bradford, *The Brussels Effect* (n 5) 37-141.

[33] Bennett and Raab (n 4) 280. See also David Vogel, *Trading up: Consumer and environmental regulation in a global economy* (Harvard University Press 1995) 5-8.

[34] Victoria Waldersee, 'Volkswagen China chief asks China's premier Li for clarity on data transfers' (2023) <https://www.reuters.com/business/autos-transportation/volkswagen-china-chief-asks-chinas-premier-li-clarity-data-transfers-2023-06-27/> accessed 20 October 2023.

[35] Commission, 'EU-China Comprehensive Agreement on Investment' (2020) <https://policy.trade.ec.europa.eu/eu-trade-relationships-country-and-region/countries-and-regions/china/eu-china-agreement_en> accessed 20 October 2023.

On the other hand, there has been no meaningful discussion on either side about China's chances of being given an adequacy decision.[37] As has been made clear in the CJEU's rulings on the adequacy decisions regarding the US,[38] the level of data protection is a matter of fundamental rights, subject to reviews of, inter alia, the powers of accessing data by the government of the third-country in question. Given the scale and intensity of digital surveillance practices in China, any consideration of the adoption of the PIPL as an attempt to gain the EU's recognition under the GDPR would be unrealistic.[39]

The partial alignment is also unlikely to be the result of any lobbying by multinational corporations. Whereas the major global powers, such as Google (Alphabet) and Facebook (Meta), have left the Chinese market, voluntarily or not, home-based technology companies breaking into the European markets have been under tight control of the state.[40] The economic drive behind the convergence between the PIPL and the GDPR is therefore much less likely to be a matter of international trade.

## 2.3. Beyond Brussels Effect

The existing frameworks are prototyped mostly in Eurocentric contexts, with a focus on the EU's unilateral actions and global power, lacking sufficient consideration of the unique context, culture, history, and interest of the 'receiving' country, such as China. They tend to demonstrate the Brussels Effect in contexts where richer and more powerful countries are in a position to set global standards.[41] However, China-specific circumstances would make the *de jure* Brussels Effect a less plausible explanation for the similarities between the PIPL and the GDPR. The challenge of applying existing theories to the case of China calls for a fresh reflection on how the GDPR interacts with the evolution of data protection laws in other jurisdictions.

The weaker *de jure* Brussels Effect may be at least partly attributed to the lack of the *de facto* Brussels Effect in China. Following Braford's reasoning, the considerable market size of the EU may mean their regulation will directly affect the practices of the businesses in a third country (*de facto* Brussels Effect), who will in turn advocate an alignment of the norms in the home country with the EU's in order to gain easier access to the European market.[42] In the case of China, however, the extent to which the GDPR directly affects Chinese citizens, corporations and even public bodies is limited. Other than the regulatory actions against TikTok,[43] a major one of which (on transborder data flows) still is pending at the time of writing, it does not seem that China is high on the EU's data protection enforcement agenda. In addition to the two points highlighted above, the lack of *de facto* Brussels Effect on them may have further disincentivised them to pursue any regulatory alignment with the EU beyond some degree of textual similarity.

---

[36] Latest statistics showed that in 2021, the EU saw a €563.6 million trade deficit in the information service sector with the US, and a €2.7 million surplus with China. Eurostat, 'International trade in services (since 2010) (BPM6)' (2023) <https://ec.europa.eu/eurostat/databrowser/bookmark/7103afb1-9e06-4017-8939-2bcb9e5a9d10?lang=en> accessed 20 October 2023.

[37] While the recent easing of China's data localisation standards appears to be a change of policy, it does not necessarily change the tide in terms of achieving adequacy. See Section 4.3 for a further discussion.

[38] Case C-362/14 *Schrems* [2015] OJ C 398/5; Case C-311/18 *Facebook Ireland and Schrems* [2020] OJ C 249/21.

[39] See Noerr, 'PIPL: Data protection made in China' (2021) <https://www.noerr.com/en/insights/pipl-data-protection-made-in-china> accessed 20 October 2023.

[40] Adam Segal, 'When China rules the web: technology in service of the state' (2018) 97 Foreign Affairs 10.

[41] Vogel (n 33) 5-8; Bennett and Raab (n 4) 280.

[42] Bradford, *The Brussels Effect* (n 5) 78.

[43] For instance, see Data Protection Commission, 'Irish Data Protection Commission announces €345 million fine of TikTok' (2023) <https://www.dataprotection.ie/en/news-media/press-releases/DPC-announces-345-million-euro-fine-of-TikTok> accessed 20 October 2023. It should be noted that, more recently, TikTok has been under investigation by the European Commission on the basis of the new Digital Services Act rather than the GDPR. See Commission, 'Commission opens formal proceedings against TikTok under the Digital Services Act' (2024) <https://ec.europa.eu/commission/presscorner/detail/en/ip_24_926> accessed 5 May 2024.

The theories examined above (the Brussels Effect included) are evidently illuminating in some cases, but they display major limitations when applied to the case of China. Overall, these theories either pivot around the EU's global regulatory or ideational powers[44] or point out the potential reconciliation between authoritarianism and the protection of privacy.[45] They do not fully explain why China, after years of struggling to find the right legal domain to protect privacy, has turned to a new regime that mirrors the GDPR to some noticeable extent. The mere fact that the European framework is (seen as) a good law is not sufficient to infer that China will have followed suit. In sum, we have three major reservations about the belief that the Brussels Effect has taken root in China in the context of data protection. First, while the EU influence is undeniably evident, it is not the only factor. The existing theories tend to emphasise the third countries' external motivation in relation to the EU, whether they have economic, political, legal or ideological roots but provide limited analysis of the country-specific factors that might significantly affect the applicability of the theories to certain countries. In the next section, we will develop the concept of 'Gravity Assist' to indicate a struggling, lengthy process of legal development in which multiple legal sectors were at play alongside the noticeable GDPR influence. Second, despite significant similarities between the GDPR and the PIPL, China has not fully embraced the GDPR as it is the gold standard. Rather, through a textual comparison we will show significant adaptations, reservations and most importantly a sweeping degree of obscurantism that creates space for discrete interpretation. This process of selective borrowing, what we refer to in this paper as 'strategic acceleration', seeks to rapidly align its own legal norms with common principles and standards with a view to gaining legitimacy and legislative efficiency. Third, it begs the question as to whether the PIPL will be enforced in a way that produces similar outcomes. We argue that it will not. Rather, there may be departures from the EU model as disagreements on the level and means of data protection stand out, particularly when balanced with other values such as economic growth and security of the party-state. More nuanced elaboration is clearly needed on how resorting to a foreign data protection regime, perceived as the 'gold standard', may diffuse domestic demands on increasing the level of data protection. To this end, we advance in the next section a more nuanced account of how the PIPL models the GDPR but retains its distinctive characters, a theory we refer to as 'Gravity Assist'.

## 3. Gravity Assist: Towards A Theory of Interactions Between Data Privacy Regimes

Gravity Assist, originating from space engineering, describes how a spacecraft utilises the gravity of another space object to facilitate the acceleration or change of course.[46] Using this term as a metaphor in the context of global technology regulation, one can describe the development of a jurisdiction's data privacy law as a move of gravity assist. It occurs when a jurisdiction approaches another more powerful regulatory regime so as to gain the necessary momentum to overcome institutional and ideological barriers and to navigate towards a different policy target (or set of targets). Particularly in the realm of data privacy, gravity assist signifies the deliberate utilisation of internal and external forces to chart a distinct course in legal development. In this paper, Gravity Assist provides a conceptual framework to understand, beyond textual similarities, the evolution of GDPR-inspired data protection law beyond Europe. This framework is devised especially to acknowledge the nuances of reluctance and resistance in local contexts, while providing a basis for evaluating its future development vis-à-vis European and international standards. Three main hypotheses are explained below, and each of them are attested in the next section. First, the development of a jurisdiction's data privacy law is a policy process in which the trajectory can be steered towards a set of internal and external regulatory domains ('*sources of*

---

[44] Bradford, *The Brussels Effect* (n 5); Bygrave, '"Strasbourg Effect"' (n 6).

[45] Jia (n 13).

[46] R. A. Broucke, 'The celestial mechanics of gravity assist' (Astrodynamics Conference, Minneapolis, August 1988).

*gravity*'). Internally, these forces can come from existing and related sectors of law that provide for path dependence, including but not limited to, criminal law, civil law, consumer law, cybersecurity law. Navigating these domains helps build up the necessary initial acceleration for the development of data privacy rules. Externally, these forces can take the form of data privacy laws in other countries, such as the GDPR. Second, Gravity Assist considers it a possibility that partial convergence of two or more jurisdictions' data privacy laws may be a strategic choice of one of those players, more likely the later entrant, taking advantage of the policymaking pull of the other country (or countries) to speed up the introduction of certain rules ('*strategic acceleration*'). In the case of the PIPL and the GDPR, our comparative analysis of the two legislative texts reveals a strong indication that the Chinese lawmakers drew on the GDPR's political appeal and technical maturity to support their adoption of the PIPL. Third, our theory also predicts that the current partial alignment of data privacy rules between certain regimes may be a temporary phenomenon. In the long term, the late-entrant country may use the regulatory capabilities gathered during the strategic acceleration to head towards a significantly different gravity source (or set of gravity sources), which will eventually capture the data privacy legal framework in a more stable manner ('*asteroid capture*'). Our analysis suggests that China's personal information protection law is likely to orbit around a set of policy goals, collectively understood as cyber sovereignty, which are in many instances not aligned with the EU data protection law, particularly in aspects other than national security.

## 4. Contextualising Gravity Assists in the Formulation of China's Legal Development

The GDPR-PIPL relationship offers an illustrative example of Gravity Assist in action. While the PIPL mirrors the GDPR in terms of structure, principles, and content to a significant extent, it strategically departs in crucial areas, carving out space for distinct interpretation and enforcement. These deviations, carefully designed to align with China's socio-legal dynamics, serve as resistance against complete capture by external legal influences. This nuanced convergence and divergence with the GDPR underscores the dynamic interplay between external influences and domestic priorities. In this section, we contextualise and substantiate the theory of Gravity Assist by contemplating the PIPL's struggling history, its nuanced mirroring of the GDPR template, and its future avenues of development in light of the Brussels Effect.

### 4.1. The Pre-PIPL Past: The Sources of Gravity

It may be reasonable to posit that, over the last three decades, China has neither completely resisted the ideas of privacy or data protection nor entirely fallen to the charms of the EU's normative power. The reluctance to recognise and ensure a high level of privacy and data protection through a stable and distinctive legal institution is apparent, as was epitomised by (1) the lack of recognition at fundamental-rights or constitutional level; (2) the absence of a freestanding statue until the PIPL; and (3) the aversion to ensure a high level of granularity of the legal provisions to ensure coherent and effective enforcement. We contextualise such a reluctance by revealing several sources of gravity within China's legal order, in interaction with the EU's regulatory global influence. It helps explain how China has come a long way, with numerous twists and turns, to legally recognise privacy and data protection as protectable interests only (rather than digital rights).

4.1.1. The Galaxy of Civil Law

The intertwined relationship between defamation and privacy marks the beginning of China's legal conceptualisation of privacy. This choice of law-making, as some scholars explain,[47] has cultural roots in China as privacy was often culturally equated with shameful secrets. As such, for almost two decades (1988-2009), privacy had been protected in China only on the condition that reputational damages were evidenced, as stipulated under Article 101 of the Civil Law General Principles (CLGP),[48] the first generation of the civil law system in China. Accordingly, privacy appears as a derivative right within the defamation law, and this narrow conceptualisation has profound ramifications for the protection of privacy and data protection.[49] After some years, efforts were then undertaken, with obvious influence from the Western privacy colleagues,[50] to draw a clearer line between reputation and privacy, which eventually contributed to a more explicit and comprehensive recognition of privacy in the Tort Liability Law 2009 (TLL).[51] Coincidentally, TLL was established at a time when the EU recognised the right to data protection as a new fundamental right next to the right to privacy. Yet, this dual footing had not been taken by Chinese legislators when designing the right to privacy along with other civil rights and interests.

---

[47] Guobin Zhu, 'The Right to Privacy: An Emerging Right in Chinese Law' (1997) 18(3) Statute Law Review 208; Tiffany Li, Jill Bronfman and Zhou Zhou, 'Saving Face: Unfolding the Screen of Chinese Privacy Law' (2016) <https://ssrn.com/abstract=2826087> accessed 20 October 2023.

The social need for privacy became increasingly pressing after the turn of the century.[52] Following almost two decades of CLGP enforcement, privacy was reconceptualised in TLL with a more explicit recognition detached from reputational torts. By nature, TLL is a specification of, and extension to, the civil law principles and concepts sketched in CLGP. It codified decades of case-law and legal scholarship and once stood as one of the earliest internet regulations with online services and platforms as its primary regulatory targets. Article 36 TLL introduces a notice and takedown mechanism that prompted China's own 'right to be forgotten' case, the *Zhu Ye* judgment.[53] Overall, TLL has a noticeably limited role in privacy protection development, partly due to the underdevelopment of privacy torts that extend to the (mis-)use of personal information and partly to the hesitation in recognising intangible harms.[54]

The Chinese civil law system has been systematically codified by 2021, with the new Civil Code (CC)[55] accommodating both the CLGP and the TLL with necessary adaptations. The making of such a mega code represented an opportunity for further developing privacy torts while considering the protection of personal information as a related but distinct protectable interest. In the CC, the right to privacy is materialised by articulating several categories of privacy torts, and personal information protection recognised for the first time as a protectable interest sitting next to the right to privacy. Despite severe debates in the legal communities,[56] China is reluctant to grant personal information protection a status of civil right[57], let alone enshrinement at constitutional or human rights levels, which contrasts significantly with the EU's fundamental-rights approach to data protection. As will be discussed further, this hierarchical arrangement has huge ramifications as personal data protection may be adjusted or overlooked in favour of broader state goals, leading to a nuanced and sometimes compromised approach to data privacy that varies with the government's strategic priorities.

---

[48] Civil Law, General Principles (China) (中华人民共和国民法通则) [1986] ('CLGP').

[49] Rebecca Ong, 'Recognition of the right to privacy on the Internet in China' (2011) 1(3) International Data Privacy Law 172.

[50] Zhu (n 47) 210.

[51] Tort Liability Law (China) (中华人民共和国侵权责任法) [2009] ('TLT').

[52] Yao-Huai Lü, 'Privacy and data privacy issues in contemporary China' (2005) 7 Ethics and Information Technology 7.

[53] Undrah B Baasanjav, Jan Fernback and Xiaoyan Pan, 'A critical discourse analysis of the human flesh search engine' (2019) 46(1-2) Media Asia 18; Ong (n 49); Dong Han, 'Search boundaries: human flesh search, privacy law, and internet regulation in China' (2018) 28(4) Asian Journal of Communication 434.

[54] Hong Xue, 'Privacy and personal data protection in China: An update for the year end 2009' (2010) 26(3) Computer Law & Security Review 284.

[55] Civil Code (China) (中华人民共和国民法典) [2020] ('CC').

[56] Quan Liu, 'Disagreements on the Rights of Personal Information Protection and Resolution (个人信息保护的权利化分歧及其化解)' (2022) 52(6) China Law Review 107; Xiaodong Ding, 'Reflection and Reconstruction of Personal Information Rights: On the Applicable Premise and Legal Interest Basis of Personal Information Protection Law (个人信息权利的反思与重塑——论个人信息保护的适用前提与法益基础)' (2020) 32(2) Peking University Law Journal 339; Fuping Gao, 'On the Purpose of Personal Information Protection: A Perspective of the Protected Interests on Personal Information (论个人信息保护的目的——以个人信息保护法益区分为核心)' (2019) 36(1) Civil and Commercial Law 93.

[57] Xinbao Zhang, 'From Privacy to Personal Information: Theory and Institutional Arrangement of Interest Rebalance (从隐私到个人信息: 利益再衡量的理论与制度安排)' [2015] Peking University Law Journal 38.

4.1.2. A Constellation of Other Laws

In parallel with the developments in civil law, China has broadened the scope of exploration by considering particular legal realms for the protection of privacy and data protection, including criminal law, consumer law, and cybersecurity law. In the meantime, China also experimented with several kinds of regulatory norms other than statutes as the basis for the ultimate enactment of the PIPL.

In the EU, data protection is primarily considered to be a non-criminal matter, despite that Article 84 GDPR provides for the potential of criminal penalties imposed upon unlawful processing activities. As De Hert and Gutwirth contend, data protection 'does not have a prohibitive nature like criminal law'.[58] Lynskey observes that Article 84 stands as a remedy to specify criminal penalties applicable for its infringement, but the decision to impose penalties is at the discretion of Member States due to the EU's lack of explicit competence to enact criminal law measures.[59] In contrast, China has taken obvious steps towards criminalising data protection matters via several amendments to its Criminal Law of 1979. These amendments make instances of 'unlawful selling', 'unlawful providing by public bodies and some critical sectors, and 'unlawful purchase and acquisition of personal information' criminally punishable if with severe consequences. As Greenleaf notes, criminal prosecutions against illegal sale or purchase of personal data were once 'the most frequently used enforcement methods' in China.[60] Yet, given the high thresholds set by the criminal law and interpretation, such criminal provisions are applied only to a limited number of cases where a high volume of data was involved and where severe consequences were caused.[61]

Consumer law was another vehicle for the protection of privacy and data protection in China. Compared to criminal law, consumer law is closer to data protection matters[62], and the US's consumer privacy approach had a considerable influence on China.[63] A consumer approach to privacy also aligns with China's political-legal landscape that, unlike the EU's omnibus data protection regime, processing of personal data in the public sector is treated distinctly. In 2012, a decision was made by China's Standing Committee of the National People's Congress[64] to improve the protection of personal information, which was at that time the one and only statute-level instrument concerning the protection of personal information. Greenleaf observes that this 12-clause decision was drafted in general terms, on which more specific regulations can be established, thereby constituting a 'gap-filling law'.[65] Indeed, since then, a range of sector-specific regulations, guidelines or standards pertaining to personal information protection were released, and many of these are of a non-binding, low-level and advisory nature. The most noticeable outcome arising from this decision was the amendments made to the consumer protection law of 1993, marking the extension of this legal framework to the digital realm. The new provisions added are, however, rather terse and characterised primarily by the introduction of informed consent. As Greenleaf puts it, it was 'almost identical with those in the […] decision' without adding more details.[66] Pernot-Leplay contends that China's approach features a distinctive data protection right placed within its consumer protection frame, coupled with several high-level principles e.g., transparency, legality and confidentiality.[67] This consumer 'right', consisting primarily of a notice-and-consent mechanism, resembles in many ways consent under the EU's Data Protection Directive.[68] The effect of this mechanism was highly in doubt as implied and *ex post* consent (or an opt out approach) were largely permitted as a way of compliance.[69]

---

[58] Paul De Hert and Serge Gutwirth, 'Data protection in the case law of Strasbourg and Luxemburg: Constitutionalisation in action' in Serge Gutwirth and others (eds), *Reinventing data protection?* (Springer 2009).

[59] Orla Lynskey, 'Article 84 Penalties' in Christopher Kuner and others (eds), *The EU General Data Protection Regulation (GDPR): A Commentary* (Oxford University Press 2020).

[60] Greenleaf, *Asian Data Privacy Laws: Trade and Human Rights Perspectives* (n 20) 225-226.

[61] Huiru Cui and Chenxing Fan, 'A Study On Judicial Application Of Crime Of Infringing Citizens' Personal Information: Based On Some Judicial Decisions In Beijing (侵犯公民个人信息罪的司法适用研究——以北京市部分司法判决为基础)' [2022](5) Journal of Shandong Police College 26.

[62] Natali Helberger, Frederik Zuiderveen Borgesius and Agustin Reyna, 'The perfect match? A closer look at the relationship between EU consumer law and data protection law' (2017) 54(5) Common Market Law Review 1427; Inge Graef, Damian Clifford and Peggy Valcke, 'Fairness and enforcement: bridging competition, data protection, and consumer law' (2018) 8(3) International Data Privacy Law 200.

[63] Pernot-Leplay (n 16).

[64] Decision of the Standing Committee of the National People's Congress on Preserving Computer Network Security (China) (全国人大常委会关于加强网络信息保护的决定) [2012].

[65] Greenleaf, *Asian Data Privacy Laws: Trade and Human Rights Perspectives* (n 20) 205.

After years of experience legislating on data privacy matters in different areas of law, it was expected that China would have been prepared for an official rollout of a comprehensive data protection law. Yet, it did not come out as expected, and the Cybersecurity Law of 2016 (CL)[70] was enacted instead. This abrupt takeover of the cybersecurity law was impacted by China's new *cyber sovereignty* agenda, first introduced in 2010 and officially announced at the 2014 World Internet Conference.[71] In China's unique context, the concept of cyber sovereignty pertains to the idea that a state should have the right to manage and control its own digital infrastructure without undue external influence to ensure security at national, ideological and economic levels.[72] It is an extension of the country's emphasis on sovereignty to the digital domain, as part of its broader push to assert its authority and influence over external influences, including the Brussels Effect.[73] Cyber-sovereignty was so influential in China that all legislative proposals that followed carried this kitemark. Far from the technical connotation concerning defence against cyber-attacks, cyber-sovereignty is broadly conceived to reflect upon, and give expression to, the national emphasis on security, stability and sovereignty.[74] Yet, the concept is elastic enough to instrumentally and pragmatically accommodate *some* data protection rules.[75] The CL is read as China 'getting one step closer' to its own GDPR[76], but the encapsulation of data protection law within an unusually broad and overstretched cybersecurity frame is contestable. Some provisions duplicate key mechanisms that can be found in the consumer protection law, and new mechanisms introduced (e.g., rights to correct, delete, breach notification etc.) evidently mimic the GDPR counterparts. That said, the CL differs from EU data protection law in significant ways. For one thing, the policy of data localisation appears much more restrictive, if not totally prohibitive, than the GDPR.[77] For another, CL is not an

---

[66] ibid.

[67] Pernot-LePlay (n 16).

[68] Directive 95/46/EC of the European Parliament and of the Council of 24 October 1995 on the protection of individuals with regard to the processing of personal data and on the free movement of such data [1995] OJ L281/31 ('Data Protection Directive').

[69] Pernot-LePlay (n 16) 211.

[70] Cybersecurity Law (China) (中华人民共和国网络安全法) [2016] ('CL').

[71] Anqi Wang, 'Cyber Sovereignty at Its Boldest: A Chinese Perspective' (2020) 16 Ohio State Technology Law Journal 395.

[72] Kuner C, 'Data Nationalism and Its Discontents' (2015) 64 Emory Law Journal Online 2089
Christakis T, 'European Digital Sovereignty: Successfully Navigating between the "Brussels Effect" and Europe's Quest for Strategic Autonomy' (2020) <https://ssrn.com/abstract=3748098>

[73] Wang (n 71).

[74] Rogier Creemers, 'China's conception of cyber sovereignty' in Dennis Broeders and Bibi van den Berg (eds), *Governing cyberspace: Behavior, power and diplomacy* (2020).

[75] Qianlan Wu, 'How has China formed its conception of the rule of law? A contextual analysis of legal instrumentalism in ROC and PRC law-making' (2017) 13(3) International Journal of Law in Context 277. See also Graham Greenleaf and Scott Livingston, 'China's New Cybersecurity Law–Also a Data Privacy Law?' (2016) <https://ssrn.com/abstract=2958658> accessed 20 October 2023.

[76] Greenleaf and Livingston (n 75).

[77] Graham Greenleaf and Scott Livingston, 'China's personal information standard: the long march to a privacy law' (2017) <https://ssrn.com/abstract=3128593> accessed 20 October 2023.

omnibus regime applicable to all sectors, with the public sector largely exempt. As Bygrave puts it, human rights are designed in the Western world mainly to protect individuals from state power, but China views these rights as derived from the state, manifesting a distinctive 'human rights philosophy that prioritises government supremacy'.[78]

### 4.1.3. A Catena of Legislative Bills

The last (internal) gravity assist takes the form of legislative bills, drafted mostly by (civil) law scholars either commissioned by state bodies or of their own accord. Unlike European specialists within governments or parliaments, it is legal scholars (primarily from the field of civil law) that play a prominent role in the law-making process in China, and the scholar-led process opens avenues to incorporate European rules and standards by way of their academic work. These 'scholarly' bills often have a direct bearing, and considerable influence over, formal legislative processes. China's legislative efforts can be traced back to the establishment of the National Information Leading Group in 2001, but the official legislative process with public consultation only commenced after 2020. In between, the first bill, commissioned by the State Council Informatisation Office in 2005 and led by Professor Hanhua Zhou,[79] had a somewhat dual legislative purpose. It concerned legislating for both personal information protection and the freedom of information, with the latter eventually ending in a separate regulation on the disclosure of government information in 2008. Zhou's bill mirrored the EU's data protection directive in some ways (e.g., principles, data subject rights, lawful basis of processing etc.) but purposefully established two different sets of rules for the public and private sectors. Ostensibly, this structural arrangement had some influence over the PIPL, for the private actors are treated differently from their public sector counterparts. The second bill of 2017, led by Congressional representative Xiaoling Wu,[80] shows contrasting characteristics. Released at the time when the GDPR was finalised, Wu's bill exhibited evident influence from the EU regulation, with the data subject rights significantly developed, notably including an explicit recognition of the right to autonomy or informational self-determination. The last bill of 2019, led by Professor Xinbao Zhang, was published at a time nearer to the official legislative process.[81] Zhang's bill characteristically put an emphasis on institutional capacity and oversight, with multiple new mechanisms introduced such as certification, expert advisory committee, corporate social responsibility, specialist organisation, government-led resource pooling and sharing mechanisms. Many of these mechanisms were, however, not eventually written into the PIPL.

### 4.2. The Present of the PIPL: Strategic Acceleration and de jure Brussels Effect in China

With a comprehensive and systematic mapping of the PIPL's past, we have shown that China had considered a wide range of legislative options ('sources of gravity') before selecting the EU model as the basis for its own legal development. The EU model prevailed *primarily* for its ideational power or regulatory capacity, assisting China in developing its own legal framework in a rapid and effective manner. Yet, the exact ways in which China takes in and adapts the GDPR template, when carefully inspected, could contest the very idea of Brussels Effect in China.

While the EU influence over PIPL is apparent, it does not fully embrace the ideas and philosophies behind EU data protection law. Existing commentaries on the PIPL tend to compare it to the EU's GDPR,[82] and one might be tempted to consider the textual similarities to be an instance of the *de jure*

---

[78] Bygrave, '"Strasbourg Effect"' (n 6).

[79] Zhou (n 31).

[80] Sohu, 'Personal Information Protection Law of People's Republic of China (Draft) 2017 (中华人民共和国个人信息保护法（草案）2017 版)' (2017) <https://www.sohu.com/a/203902011_500652> accessed 20 October 2023.

[81] Zhang (n 31).

[82] For a comparison conducted by practising lawyers, see Albrecht (2021), Deng and Dai (2021), Ke et al. (2021), Laird (2022), Potter (2022), Zhang (2022), Zhu (2021); For those by academics, see Sheng and Tang

Brussels Effect. This subsection offers a comparative analysis of the two legislative texts concerned, informed by our best efforts to assemble scattered evidence of legislative history and debate - a daunting task owing to the restricted public access to the legislative process. We will qualify the nature of the areas of convergence and divergence between the two laws, highlighting a significant degree of adaptations, reservations and, most importantly, a sweeping degree of obscurantism that creates space for discretionary interpretation. It thereby casts some doubts on the extent to which the (*de jure*) Brussels Effect holds true in China. This section is neither intended to repeat the existing analyses nor to conduct an in-depth, complete comparison between the two legal frameworks. We will unearth the subtle differences between the two regimes that are understated in the literature, debuting some claims that differentiate these two regimes on contested grounds. When situated within China's own legislative contexts, the GDPR is, in our view, but one of the templates that China utilises to legitimise and rationalise its own legal framework. This process of selective borrowing marks a phase of 'strategic acceleration' in which the Chinese law rapidly gains legislative efficiency and legitimacy while aligning its own legal norms with common principles and standards.

4.2.1. Areas of convergence

A number of key components of the PIPL show clear alignment with the GDPR, and in some cases, also its predecessor, the Data Protection Directive. Some of these aspects are very unique to the GDPR at the time when the PIPL was enacted, and therefore, it would be rather unlikely that such alignments were a pure coincidence between parallel systems.[83]

The first area is a clear alignment between the two laws in their *scope*. In terms of territorial scope, Article 3 PIPL sets out three scenarios where the law applies extraterritorially: (a) 'provide products or services to natural persons within the borders'; (b) 'analysing or assessing activities of natural persons within the borders'; and (c) 'other circumstances provided in laws or administrative regulations'. The first two scenarios are noticeably similar to those set out in Article 3(2) GDPR, i.e. 'offering of goods or services […] in the Union' and 'monitoring of their behaviour […] within the Union'. In terms of material scope, Article 72 PIPL provides a clear exclusion of its applicability to 'natural persons handling personal information for personal or domestic matters', significantly similar to the household exemption in Article 2(2)(c) GDPR, i.e., 'a natural person in the course of a purely personal or

---

(2022), Solove (2021). For a comparison conducted by legal practitioners, see Daniel Albrecht, 'New Developments in Chinese Data Protection Law in Contrast to the European GDPR—New requirements for network operators in China' (2021) 22(5) Computer Law Review International 142; Jet Deng and Ken Dai, 'The Comparison Between China's PIPL And EU's GDPR: Practitioners' Perspective' (2021) <https://www.mondaq.com/china/data-protection/1122748/the-comparison-between-chinas-pipl-and-eus-gdpr-practitioners-perspective> accessed 20 October 2023; Ke and others (n 12); Catherine Zhu, 'Is China's New Personal Information Privacy Law the New GDPR?' (2021) <https://news.bloomberglaw.com/privacy-and-data-security/is-chinas-new-personal-information-privacy-law-the-new-gdpr> accessed 20 October 2023; Jennifer Laird, 'The GDPR vs China's PIPL' (2022) <https://www.privacypolicies.com/blog/gdpr-vs-pipl/> accessed 20 October 2023; Angela Potter, Keshawna Campbell and Victoria Ashcroft, 'Comparing Privacy Laws: GDPR v. PIPL' (2022) <https://www.dataguidance.com/resource/comparing-privacy-laws-gdpr-v-pipl> accessed 20 October 2023; Thomas Zhang, 'GDPR Versus PIPL – Key Differences and Implications for Compliance in China' (2022) <https://www.china-briefing.com/news/pipl-vs-gdpr-key-differences-and-implications-for-compliance-in-china> accessed 20 October 2023. For those by academics, see Daniel Solove, 'China's PIPL vs. the GDPR: A Comparison' (2021) <https://teachprivacy.com/chinas-pipl-vs-gdpr-a-comparison> accessed 23 October 2023; Xiaoping Sheng and Junjie Tang, 'A Comparative Analysis of Personal Information Rights in China and Personal Data Rights in EU: Based on the Personal Information Protection Law and GDPR (我国个人信息权利与欧盟个人数据权利的比较分析：基于《个人信息保护法》与 GDPR)' (2022) 66(6) Library and Information Service 26.

[83] For instance, the references to the extraterritorial scope of the law appeared for the first time in the 2012 Commission proposal for the GDPR, and in the case of the PIPL, the legislative draft during the first reading in 2020. Prior to that, in Chinese legislation we are unable to find any similar reference to phrases like 'provide products or services to natural persons within the borders' or 'analysing or assessing activities of natural persons'. For these reasons, the divergence is much more likely to be an influence of the GDPR rather than an organic development from the domestic legal order.

household activity'. While the household exemption is not unique to the GDPR but rather a reiteration of the equivalent in the DPD that can find its origin in the 1982 Swedish Data Act,[84] the extraterritorial provisions can be clearly attributed to the GDPR, rendering it fair to assume that the PIPL found inspiration from the GDPR in this respect.

The second area of convergence concerns some similar *definitions and concepts* under the two laws. For instance, the definitions of personal information in Article 4 PIPL and personal data in Article 4(1) GDPR are both developed around the key concept of 'identified or identifiable natural person'. It should be clarified that the sheer textual similarity does not mean the same concept will be interpreted and enforced the same way. For example, the privacy scholarship in China has made the argument that the exact scope of personal information should be determined in relation to the risks involved,[85] which may have an impact on how the definition will be understood in practice. However, this is not reflected in the text of the PIPL, and it remains unanswered why the legislators chose such a similar terminological approach despite potential underlying conceptual differences. A case in point for comparison is the determination of the nature of vehicle registration number. Both jurisdictions have adjudicated on this matter with similar conclusion - that vehicle registration number does not constitute personal data in principle. Whereas the Guangzhou Internet Court alludes to the low likelihood of using such data alone to identify individuals and upon the difficulties in combining with other sources data in reality,[86] the CJEU adheres to the broad threshold of 'reasonably likely' established in *Breyer*.[87] The concept of joint information handlers under Article 20 PIPL, or in the GDPR's terms, joint data controllers, has also shown some clear mirroring with Article 26 GDPR in terms of both the definition and the legal consequences. Other concepts, such as automated decision-making (Articles 24 and 73(2) PIPL; Article 22 GDPR), sensitive personal information (Article 28 PIPL; 'special categories of personal data', Article 9 GDPR) and de-identification (Article 73(3) PIPL; 'pseudonymisation', Article 4(5) GDPR).[88] Most of these concepts are inherited from the DPD but it would be fair to assert that they mirror their equivalent in the European regime.

The third area primarily concerns the key *principles* of the legal frameworks. Articles 5 to 9 PIPL lays down a set of general principles governing the handling of personal information. Unlike the seven principles provided for in Article 5 GDPR, the PIPL has not given specific names to those principles. Despite the differences in the exact wording, the substance of these provisions can be loosely compared to the GDPR principles of 'lawfulness, fairness and transparency' (Articles 5, 7 PIPL; Article 5(1)(a) GDPR), 'purpose limitation' (Article 6 PIPL; Article 5(1)(b) GDPR), 'data minimisation' (Article 6 PIPL; Article 5(1)(c) GDPR), 'accuracy' (Article 8 PIPL; Article 5(1)(d) GDPR) and 'accountability' (Article 9 PIPL; Article 5(2) GDPR). The principle of storage limitation (Article 5(1)(e) GDPR) can also be found in Article 19 PIPL. It should however be noted that these principles are not unique to the GDPR or its predecessor; quite the contrary, many of these principles can be traced back to the 1980

---

[84] Jiahong Chen and others, 'Who Is Responsible for Data Processing in Smart Homes? Reconsidering Joint Controllership and the Household Exemption' (2020) 10(4) International Data Privacy Law 279.

[85] Aimin Qi and Zhe Zhang, 'Identification and reidentification: The definition of personal information and the legislative choice (识别与再识别: 个人信息的概念界定与立法选择)' (2018) 24(2) Journal of Chongqing University (Social Science Edition) 119; Li Zhang and Duoqi Xu, 'On the legal regulation of personal information anonymizationin China under the risk control concept (风险控制理念下我国个人信息匿名化处理的法律规制)' (2023) 29(2) Journal of Chongqing University (Social Science Edition) 220.

[86] *Yu v Chaboshi (余某某诉查博士)* [2021] 粤 0192 民初 928 号.

[87] Case C-319/22 *Gesamtverband Autoteile-Handel (Accès aux informations sur les véhicules)* [2023].

[88] There are conceptual disputes as to how the concept of de-identification overlaps with that of pseudonymisation. For instance, see Jules Polonetsky, Omer Tene and Kelsey Finch, 'Shades of gray: Seeing the full spectrum of practical data de-intentification' (2016) 56 Santa Clara Law Review 593. In recognition of such conceptual issues, we contend that these concepts can be used interchangeably in this specific comparative context.

OECD Privacy Guidelines[89] or the 1981 Council of Europe Convention 108.[90] Regardless of whether the legislators of the PIPL took inspiration from the GDPR or elsewhere, convergence of the two regimes in this respect is evident. Further, some principles can be said to be unique to China, such as the principle of good faith (Article 5 PIPL). The principle seems to be a mere reiteration of what has always been an overarching doctrine in Chinese civil law (see Article 7 CC). The issue of how this principle operates in the context of data privacy and how it interacts with other PIPL principles (e.g., fairness) are matters worth investigating in future research.

As a fourth area of convergence, the PIPL and the GDPR also share some important *safeguards* to protect individuals, including rights given to data subjects (information subjects) and duties imposed on data controllers (information handlers). On the data subject's side, consent and transparency are perhaps the most prominent mechanisms designed to empower individuals. The alignment between the two laws in this respect can be seen with regard to the conditions for valid consent set out in the PIPL and the GDPR. Three out of the five new conditions introduced by the GDPR, i.e. right to withdraw (Article 15 PIPL; Article 7(3) GDPR), conditionality (Article 16 PIPL; Article 7(4) GDPR) and parental consent (Article 31 PIPL; Article 8 GDPR) can find their counterparts in the PIPL.[91] Similarly, the GDPR rights such as the right to information (Article 17 PIPL; Article 13 GDPR), right of access (Article 45 PIPL; Article 15 GDPR), right to rectification (Article 46 PIPL; Article 16 GDPR), right to erasure (Article 47 PIPL; Article 17 GDPR) and right to data portability (Article 45 PIPL; Article 20 GDPR) can also be found in PIPL with a varying degree of adaptations. With these subject-side mechanisms, data protection standards in China appear akin to those under the GDPR. On the data controller's side, certain procedural and organisational measures are also required by both the PIPL and the GDPR, such as the appointment of data protection officers (Article 52 PIPL; Articles 37-39 GDPR), impact assessment (Article 55 PIPL; Article 35 GDPR) and breach notification (Article 57 PIPL; Articles 33-34 GDPR). In sum, our comparative analysis suggests a sophisticated convergence exists between the two legislations across various facets. Inferred from the synchronicity of their enactment, historical development and evident legislative inclinations, these similarities are unlikely to be coincidental. Instead, they reveal a tangible linkage between the two legislations, though the practical enforcement and interpretative nuances might diverge, as will be argued later, owing to China's unique administrative and regulatory landscape.

4.2.2. Areas of divergence

Despite all the similarities between the PIPL and the GDPR, major differences exist in some of the most important areas. One seemingly apparent divergence is that the PIPL has a specific section dealing with personal information handled by state authorities. This may come across as an indication that, unlike the GDPR, the PIPL is less of an omnibus framework that covers both public- and private-sector uses of personal data. However, a closer look at Articles 34-37 PIPL would make it clear that these provisions only address such matters as the requirements for a statutory basis (Article 34), exemptions from the duty of notification (Article 35) and local storage of information (Article 36). These are, on paper, not much different from how the GDPR regulates the processing of personal data on the basis of Article 6(1)(e) ('the performance of a task carried out in the public interest or in the exercise of official authority', see also Articles 6(2), 6(3), 23, 49(3) GDPR). The degree of divergence in terms of the two regimes' coverage of the public sector is hence not as stark as it might seem at first glance.

---

[89] OECD, 'OECD Guidelines on the Protection of Privacy and Transborder Flows of Personal Data' (1980) <https://read.oecd-ilibrary.org/science-and-technology/oecd-guidelines-on-the-protection-of-privacy-and-transborder-flows-of-personal-data_9789264196391-en> accessed 13 June 2020.

[90] Council of Europe Convention for the Protection of Individuals with regard to Automatic Processing of Personal Data [1981] ETS No. 108.

[91] There are however subtle differences in the terminology that might have significant practical implications. For example, when it comes to the use of sensitive data based on consent, the GDPR requires the consent to be 'explicit' while the PIPL standard is 'separate consent'. Neither law has defined these terms, which leaves these concepts open to interpretations to data protection authorities and courts.

The more fundamental difference perhaps lies in the *institutional arrangements* to enforce the law. Under the GDPR, the agencies overseeing the application of data protection law are the independent supervisory authorities in each Member State.[92] At the Union level, the European Data Protection Board (EDPB) is set up to ensure consistent implementation of the law and to facilitate cooperation.[93] The independent status of the national supervisory authority and the EDPB is not only a legal requirement under the GDPR[94] but indeed a 'constitutional guarantee' under Article 8(3) of the Charter of Fundamental Rights.[95] The PIPL, on the other hand, assigns such functions to central and local governmental departments.[96] At national level, the responsible authority is the Cyberspace Administration of China (CAC), which is also the same organisation of the Office for Cybersecurity and Informatisation of the Chinese Communist Party. Yet, various other state bodies are playing a certain role in the enforcement of the PIPL, including the Ministry of Industry and Information Technology (MIIT), which has been put in charge of the audits of (mobile) applications, and State Administration of Market Regulation (SAMR), who might intervene from a consumer and competition angle. Sector-specific data protection matters are also subject to the oversight of the relevant department bodies such as the People's Bank of China.[97] These arrangements show that the enforcement of the PIPL follows a much less centralised model than the EU's.[98] Such a difference is likely a result of the multitude of sources of gravity highlighted in Section 4.1 above, with, for example, the pre-PIPL legacy of using consumer law enforcement mechanisms to protect consumer data remaining relevant.

Another difference between the PIPL and the GDPR is the former's emphasis on *ex ante* state oversight when it comes to data localisation requirements. As a matter of general rule, data transfers to outside China must satisfy one of the conditions set out in Article 38: (a) security assessment organised by the CAC; (b) personal information protection certification under rules set out by the CAC; (c) a standard contract formulated by the CAC; or (d) other bases provided in laws or administrative regulations or by the CAC. While conditions (b) and (c) above can find their equivalent in Article 46 GDPR, which sets out the 'appropriate safeguards' for extra-EU data transfers,[99] the options provided by the PIPL are much narrower and clearly more centred around the role of the CAC. Unlike the GDPR, the PIPL does not provide for a set of 'derogations for specific situations',[100] such as consent, where the appropriate safeguards are not in place. Controversially, consent is under the PIPL an additional, accumulative requirement to the four conditions mentioned above,[101] and cannot legitimate international data transfers alone. It should be noted, however, that recent regulatory adjustments in China have seen a slight easing of these stringent measures, as detailed in the CAC's new measures that allow for more flexibility in the cross-border transfer of data.[102] While some operational constraints may have been lessened from March 2024, the necessity for state-centric procedures continues to indicate a fundamentally different approach and therefore an enduring divergence in how personal data is governed across borders.

---

[92] GDPR, ch VI.

[93] ibid ch VII.

[94] ibid arts 52, 69.

[95] Charter of Fundamental Rights of the European Union [2012] OJ C326/391 ('Charter').

[96] PIPL, art 60.

[97] The People's Bank of China, 'Notice of the People's Bank of China on the Public Consultation on Banking Sector Data Security Management Measures (Consultation Draft) (中国人民银行关于《中国人民银行业务领域数据安全管理办法（征求意见稿）》公开征求意见的通知)' (2023) <www.pbc.gov.cn/tiaofasi/144941/144979/3941920/4993510/index.html> accessed 20 October 2023.

[98] See Yaping Gao and Jing Xu, 'Unclear Supervisors Behind App Personal Information Protection' (2022) <https://www.dehenglaw.com/CN/tansuocontent/0008/025360/7.aspx?MID=0902> accessed 20 October 2023.

[99] GDPR, art 46(2)(c), (f) ('standard data protection clauses' and 'approved certification mechanism').

[100] ibid art 49.

[101] PIPL, art 39.

[102] Measures for Promoting and Regulating Data Cross-border Transfers (China) (促进和规范数据跨境流动规定) [2024].

While the PIPL has not specified the factors and criteria for a security assessment, the Measures for Data Export Security Assessment, issued by the CAC in May 2022, makes it clear that such assessments should 'in particular assess the risks of data export activities on national security, public interest, and the legitimate rights and interests of individuals or organisations'.[103] Such a focus on national security is perhaps not a surprise given one of the PIPL's principles being the prohibition on 'personal information handling activities harming national security or the public interest'.[104] In this regard, compared to the GDPR's roles in facilitating intra-EU/EEA data flows and ensuring continuity of protection for extra-EU/EEA data transfers, the PIPL's data localisation requirements have a much stronger focus on safeguarding national security. Having the aim of safeguarding national security in data privacy law is, however, not unique to China. Finland's Personal Data Registers Act 1987, for instance, made specific reference to ensuring the security of the State, although this was removed when the 1987 Act was replaced by the Personal Data Act 1999.[105] That said, with the combination of the ministerial enforcement model and the tight control on cross-border data flows, the level of focus on security is *arguably exceptional*.[106] Again, the legacy of internal sources of gravity – in this case, the Cybersecurity Law as discussed above – are likely to have caused such a divergence.

In sum, these two areas of divergence do not represent all dissimilarities between the GDPR and the PIPL but signal what might be fundamental differences in the policy goals of the two instruments. In the foreseeable future, they might in turn drive the data protection regime in China towards different directions from what Europeans seek to achieve. Moreover, none of the four areas of convergence highlighted above would be incompatible with the strategic objective inferable from the two areas of divergence, namely the national security interest of the state. The choice of the aligned and dis-aligned components in the PIPL can be seen as a move by China's policymakers to use the GDPR instrumentally and pragmatically as a template for legislation, mirroring the GDPR's basic structure, principles, and major mechanisms but in the meantime creating space for distinct interpretation and enforcement. When bundled in one single piece of legislation, the areas of convergence can provide the normative legitimacy and technical maturity that would boost the growth of China's regime.

One technique to ensure the policy goal will indeed benefit from, rather than be impeded by, the acceleration is to obscure and reverse-detail various provisions that the GDPR was intended to specify and elaborate, particularly those hardly alignable with Chinese ideologies, politics and cultures. For example, both Article 19 PIPL and Article 5(1)(e) set out an exception to the storage limitation principle, but the scope of the former is much wider, allowing any administrative regulations to create an exemption. The *de jure* Brussels Effect, to such an extent, is at best a phenomenon on paper, and at worst a misconceptualisation of the normative dynamics between the EU's data protection regime and the one of China. What appears to be a regulatory convergence may turn out to be a temporary, strategic move as part of a broader data privacy agenda hijack that might eventually give rise to global regulatory divergence. In this regard, there is a fundamental functional departure of the PIPL from the GDPR, which reflects what Creemers would call 'the contrast in security concepts between Beijing and its Western counterparts',[107] calling into question the conceptual validity of using Western understandings of data protection law to approach the PIPL.[108]

### 4.3. The Post-PIPL Future: Asteroid Capture and Cyber Sovereignty

---

[103] Measures for Data Export Security Assessment (China) (数据出境安全评估办法) [2022], art 8.
[104] PIPL, art 10.
[105] Lee A Bygrave, *Data Privacy Law: An International Perspective* (Oxford University Press 2014) 120.
[106] For a similar viewpoint, see Graham Greenleaf, 'China's Completed Personal Information Protection Law: Rights Plus Cyber-security' (2021) 20-23 <https://dx.doi.org/10.2139/ssrn.3989775> accessed 20 October 2023.
[107] Creemers, 'China's conception of cyber sovereignty' (n 74).
[108] For a broader point about this regarding the Chinese legal system, see Donald Clarke, 'Anti anti-Orientalism, or is Chinese law different?' (2020) 68(1) The American Journal of Comparative Law 55.

In this subsection, we consider future trajectories of personal information protection development in China in the light of Gravity Assists. As previously indicated, numerous forms of gravity assists within China's legal order have contributed to the maturation of personal information protection in China. These gravity assists did not, however, eventually bring clarity to the relationship between PIPA and other related areas of law. In the absence of functional human-rights or constitutional scrutiny, it is also elusive how personal information protection as a protectable interest could be balanced against other legal rights and interests, together with broader social imperatives such as economic growth and cyber sovereignty.

The GDPR, as an external form of gravity assist, certainly has a significant and explicit shaping effect on China's PIPL. The comprehensive, omnibus, rights-based model it represents has prevailed over other modes of regulation (e.g., US consumer and sectoral approach) and laid a foundation for the current PIPL. The Brussels Effect in terms of structural and textual resemblances of the legal provisions is, however, controversial. Admittedly, such resemblances might eventually lead to some degree of convergence in the areas of consent, principles, and impact assessments, but the enforcement strategies and priorities, we contend, would be considerably different in China.

In the mid and long terms, personal information protection is by nature somewhat different from what the EU holds dear about the fundamental right to data protection. As previously presented, their fundamental differences can be explained in part from the ways in which the GDPR provisions are *selectively* borrowed and in part from how personal information protection is contextualised and institutionalised in China's legal order. The latter is relatively unspoken, and this sub-section intends to provide a detailed account of it. In our view, the influence of cyber sovereignty is overwhelmingly influential to the extent that it fundamentally alters the nature of personal information protection. It does so by dissolving the foundational premises of data protection that are seen as intrinsic and foundational to Europeans, thereby rendering the EU's influence interim, contingent and ephemeral. To this end, we introduce the concept of 'asteroid capture' to indicate that China's deeply entrenched national strategy is bound to offset the EU's influence at a certain stage, and will predominantly determine the PIPL's future trajectories. This would happen as if a larger planetary body (i.e., cyber sovereignty) eventually and permanently captures an asteroid (i.e. the PIPL) as its natural satellite. In this subsection, we present how information protection law may be understood, interpreted and enforced in a way that differs from the EU, displaying many levels of resistance to the GDPR and their underlying rationale. Three main observations are made, concerning respectively the PIPL's under-privileged status in China's legal order, its subtle departure from a rights-based, omnibus regime, and its inextricable link to other areas of security law (including Cybersecurity Law (CL), National Security Law (NSL), and Data Security Law (DSL)[109]).

First, unlike the EU's cherishment of data protection as a freestanding, fundamental right, China's placement of personal information protection as a legally protectable interest in its legal order is further down the legal hierarchy to such an extent that its legal position has significantly affected how personal information protection is balanced against other rights and interests. Particularly relevant rights include the right to conduct a business (not formulated explicitly in China), economic growth, as well as cyber sovereignty. Personal information protection, even combined with the right to privacy, has never been elevated to the highest level of hierarchy as a fundamental or constitutional right in China. There is ample evidence indicative of its secondary nature. For instance, the question of whether a (civil) rights status should be assigned to personal information protection had been fiercely debated in the legislative process.[110] While personal information protection was recognised in the CC as a freestanding legal

---

[109] CL; National Security Law (China) (中华人民共和国国家安全法) [2015] ('NSL'); Data Security Law (China) (中华人民共和国数据安全法) [2021] ('DSL').

[110] Liu (n 56); Weixing Shen, 'On the Construction and Systematization of the Personal Information Right (论个人信息权的构建及其体系化)' (2021) 189(5) Journal of Comparative Law 1; Xinbao Zhang, 'The Structure of Rights and Interests Relating to Personal Information (论个人信息权益的构造)' (2021) 33(5) Peking University Law Journal 1144; Xixin Wang, 'The Three-level Frame and Protection Mechanism of Personal

interest next to the right to privacy, it was not assigned a right status, presumably rendered less protected than other civil rights. Moreover, a section on information subject rights is indeed added to the PIPL mirroring the GDPR's data subject rights, notably including data portability,[111] but the provisions are largely obscured, the granularity of which resembles somewhat to Article 12 of the EU's Data Protection Directive. Much space is left for local courts to decide on the nature, scope and impact of these rights on a case-by-case basis. It should be noted, as a significant caveat, that there exists a scant corpus of case-law, and the few decisions that are publicly accessible are inadequate to form a coherent pattern. That said, it suffices to say that judicial interpretations primarily transpire within the private sector as actions against public entities are infrequent. Moreover, courts appear to be in a more tolerant stance when balancing individual rights against broader economic imperatives and business freedoms.[112] This inclination stems in part from the diminished granularity of the provisions and in part from deficient theoretical understanding of the potential harms arising from data processing activities, which further complicates judicial reasoning.

Second, China's PIPL differentiates itself from the GDPR prototype by departing from a rights-based, omnibus regime, which Lynskey summarises as 'key characteristics of EU data protection law'.[113] We argue this in recognition of the fact that the PIPL does include a section on information subject rights and does in theory indiscriminately apply to the public sector as well. Yet, such superficial resemblances do not reflect, but would potentially mislead, the perception of the evolution of China's PIPL. Even though China has subscribed to the EU regulation's structure, content, and underlying ideas, it has significantly altered the nature and working of data protection, through tweaks and reservations in the borrowing process, in a way that prefers state action over individual empowerment, sectoral over omnibus regime. The system of information protection rights put in place, for instance, remains barely operationalised, with little or no guidance from the regulator. Despite the interpretative powers granted by the PIPL to the CAC, the authority remains relatively passive in elucidating and detailing the provisions of the PIPL through official guidelines. Consequently, it is in practice the courts that, on an ad hoc basis, step in to interpret and clarify these largely obscure provisions, occasionally drawing on EU experiences to inform their decisions. The recent *Vipshop* judgment, for instance, represents a good attempt to clarify by court the scope, nature and conditions of the right of access.[114] Additionally, the high-profile judgment of *Hangzhou Safari Park* on the private use of facial recognition arguably concerns a request to delete personal information but was not made on the basis of the PIPL due to the timing of the litigation prior to the PIPL's taking effect.[115] The judgments delivered by local courts barely reach higher-level adjudication to exert a national-level influence, and their scope is largely determined by the individual cases. Their influence might increase when selected by the Supreme Court as exemplars[116], yet such precedent-based guidance is rather contextual, inadequate and potentially inconsistent. As of yet, it is far from conclusive that the exercise of information protection rights can be reliably facilitated through siloed judgments of a local impact. It is observed that, due to the regulatory preference for centralised, authoritative measures over empowering individuals with civil liberties and freedoms, China's authorities favour a public governance approach characterised by the

---

Information Rights and Interests (个人信息权益的三层构造及保护机制)' (2021) 43(5) Modern Law Science 105.

[111] PIPL, art 45; GDPR, art 20.

[112] See for instance, *Ling v Tiktok Technology Co., Ltd (凌某某诉北京微播视界科技有限公司隐私权、个人信息权益网络侵权责任纠纷案)* [2019] 京 0491 民初 6694 号; and *Huang v Tencent (黄某诉腾讯科技（深圳）有限公司广州分公司、腾讯科技（北京）有限公司隐私权、个人信息权益网络侵权责任纠纷案)* [2019] 京 0491 民初 16142 号.

[113] Orla Lynskey, *The Foundations of EU Data Protection Law* (Oxford University Press 2015) 14-40.

[114] *Zhou v Vipshop (周彦聪诉广州唯品会电子商务有限公司个人信息保护纠纷案)* [2022] 粤 01 民终 3937 号.

[115] *Guo v Hangzhou Safari Park (郭兵诉杭州野生动物世界有限公司服务合同纠纷案)* [2021] 浙 01 民终 10940 号.

[116] See Stanford Law School, 'China Guiding Cases Project' (2021) <https://law.stanford.edu/china-guiding-cases-project> accessed 20 October 2023.

focus on non-compliance misbehaviours (e.g. excessive collection of personal data, use beyond defined purposes, illegal selling of personal data) rather than empowerment and remedies. A prime illustration of this favoured approach is the MIIT's primary work on mobile applications inspections, with a focus on illegal use of third-party software development kits (SDKs)[117], which contrast the lacuna of enforcement or guidance from several competent authorities.

Third, somewhat reminiscing of the abrupt takeover of 'cybersecurity' in 2016, the trajectory of personal information protection is undetachable from China's security agenda, marking personal information protection as a highly politicised realm. This has far-reaching and multi-faceted implications for how personal information protection is understood, interpreted and enforced. Creemers argues that there are a total of three competing objectives for China's personal information protection.[118] In addition to typical ones shared with the EU, i.e., the protection of data protection and free flow of personal data, 'cybersecurity' broadly understood stands as a third independent objective that co-shapes the nature and enforcement of the PIPL. Contrary to how this legislation is named, the latter objective seems to stand as a more pressing imperative than the other two to the extent that PIPL may serve the interests of the party-state much more than the achievement of a high and consistent level of personal information protection. Indeed, the emphasis on cybersecurity seems conceptually fuzzy and has created difficulties in maintaining coherence within China's legal order. Cybersecurity generally refers to the practice of protecting systems, networks and programmes from attack but in China's political climate, this is not just a technical issue but a strategic one crucial to national security, involving comprehensive control measures and state oversight over cyber infrastructure and the flow of information.[119] As Creemers puts it, China has sought to 'institute and consolidate effective control over online actors, activities, and content through a process of territorialization, indigenization, and investment, while maintaining technical interoperability with the global Internet'.[120] Hence, rather than being a tool that seeks to maintain a high level of data protection understood as a functional and active exercise of data rights and accountable processing practices, personal information protection in China is better viewed as a tool for social governance. Without conceptual clarity, there are three freestanding pieces of legislation in relation to personal data with apparent overlaps and shared objectives. Whereas the PIPL is devised to primarily maintain a good level of data protection, the DSL deals primarily with technical matters against cyberattacks. The preceding Cybersecurity Law, on the other hand, appears to be an overarching scheme that speaks immediately to cyber sovereignty, with some of the provisions outdated due to the enactment of the PIPL and the DSL. The imposition of fines on Didi (a taxi-hiring app),[121] and recent amendments to harmonise the three pieces of legislation[122] are a vivid illustration of such overlaps. Unlike the EU authorities that proactively initiate actions, there has been only one primary imposition of fines on Didi, which is not even wholly based upon the PIPL. China exhibits a very different logic and strategy of enforcement to ensure that cyber sovereignty in the local context is respected. Further, China has a global trade aspect that involves, primarily, the trade war with the United States that is primarily manifested by the battle between supranational technology companies. A severe command-and-control approach would significantly impede the competitiveness of China's companies vis-a-vis foreign rivals and as such, cyber sovereignty dictates that enforcement be selective, deterrent and sometimes only declaratory. The recent easing of data localisation requirements in China might appear a shift away from piroitising security, thus aligning more closely to the EU's approach.[123] We argue that this is oversimplistic. Actually, the refined data flow policies align naturally with China's broader diplomatic policy (i.e. Digital Silk Road') on the one hand[124], and with its urgent need for a post-pandemic economic recovery and transformation on the other (through building a data marketplace to boost economic recovery), thus serving as another vivid example of selective alignment.

---

[117] MIIT, 'Notice on APPs (SDK) Infringing User Interest (关于侵害用户权益行为的 APP（SDK）通报（2024 年第 2 批，总第 37 批）)' (2024) <https://wap.miit.gov.cn/xwdt/gxdt/sjdt/art/2024/art_424e1faa0479457689f7a49e113a5429.html> accessed 5 May 2024.

[118] Creemers, 'China's conception of cyber sovereignty' (n 74).

[119] Austin G, 'International Legal Norms in Cyberspace: Evolution of China's National Security Motivations' in Anna-Maria Osula and Henry Rõigas (eds), International Cyber Norms: Legal, Policy & Industry Perspectives (NATO CCD COE Publications 2016)

[120] Creemers R, 'China's Conception of Cyber Sovereignty: Rhetoric and Realization' in Dennis Broeders and Bibi van den Berg (eds), Governing Cyberspace: Behaviour, Power and Diplomacy (Rowman & Littlefield 2020) 109-110.

[121] Gabriela Kennedy and Joshua T. K. Woo, 'The CAC is Coming: Didi Chuxing Fined a Record-breaking USD 1.2 Billion for Breach of Data Protection Regulations' (2022) <https://www.mayerbrown.com/en/perspectives-events/publications/2022/08/the-cac-is-comingdidi-chuxing-fined-a-recordbreaking-usd-12-billion-for-breach-of-data-protection-regulations> accessed 20 October 2023.

[122] Reuters, 'China looks to increase penalties under its cybersecurity law' (2022) <https://www.reuters.com/world/china/china-seeks-public-comment-possible-amendments-cybersecurity-law-2022-09-14/> accessed 20 October 2023.

## 5. Conclusion

The evolution of China's PIPL is inherently path-dependent but defies facile categorisation within existing theoretical frameworks like the Brussels Effect, raising conceptual challenges for understanding its origin, motivations, and actual level of alignment with European and international standards. While multiple elements of the PIPL mirror the GDPR, the underlying dynamics that shape China's legislative trajectory, as previously revealed, are far from linear and one-dimensional. We introduce Gravity Assist as a conceptual framework to hence provide a novel and nuanced lens to capture and perceive the intricacies of China's data protection journey. China's legislative orbit is influenced, we argue, not only by the gravitational pull of the GDPR, but local contextual factors that significantly and determinatively alter its trajectory now and then. Particularly, we provide a detailed analysis of how the overarching strategy of cyber sovereignty has interrupted its original trajectory shaped by the GDPR and will continuously and significantly dictate its future trajectories.

The Brussels Effect, while enlightening, tends to fall short of engaging local contexts and hence requires additional contextualisation to encapsulate the entirety and complexity of the lawmaking process in other non-Western jurisdictions beyond mere legal borrowing. To better understand the evolution of China's GDPR-inspired data protection law under the influence of the EU, it is critical to turn the perspective inside out and attend to how China pulls and takes in, rather than receives, GDPR-equivalent standards. Flipping the narrative would give us more nuances and clarity as to how these norms are understood, enforced and where relevant, altered. It is on this basis we develop the theory of Gravity Assist to enrich our understanding of data protection development in several ways. First, it unveils China's strategic adoption of the GDPR as a multi-faceted process, involving not just textual alignment but also intricate policy considerations. We place this process with historical and policy contexts, highlighting how China's legislative journey involves several waves of legislative attempts, challenges and adaptations. Framing the GDPR as one of the many 'sources of gravity' that propel China's data protection law forward, we gain insights into why such alignment was needed, how it fits into China's broader digital strategy, and how it might evolve in a way that derails from the EU model in the future. It is revealed that the relationship between the GDPR and the PIPL is not a one-dimensional replication but a dynamic interplay between external influences and domestic priorities. The theory of Gravity Assist hence offers a more comprehensive and country-specific perspective on the complexity of data protection evolution, especially in non-Western jurisdictions, extending beyond the assumptions of the Brussels Effect. Second, the theory of gravity assists elucidates the nuanced relationship between the PIPL and the GDPR despite apparent similarities in content, form and structure. Countries where cultural values and legal traditions diverge from the privacy-centric ethos of Western counterparts often lack the intrinsic incentive and power to autonomously propel itself into a robust data protection framework. Here, external gravitational forces, represented by established data protection laws like the GDPR, provide the necessary momentum for change. From the non-EU-centric perspective, this also explains the EU's ideational power, as indicated by Bygrave,[125] that presumably plays an even more significant role, particularly at the early time of legislative consideration. Gravity Assist thus becomes a natural and necessary process through which legal systems incrementally familiarise with, adapt to, test with, and ultimately incorporate data protection principles into their legal fabric. In light of the numerous adaptations, reservations and the overall generalisation, it is concluded that the GDPR was not wholly embraced in China as the gold standard despite noticeable similarities between the GDPR and the PIPL. Third, in addition to explaining the coming into being of China's PIPL, the theory also addresses future developments of China's PIPL and cross-jurisdictional regulatory trajectories at global and sub-global levels, taking into account the interpretation, enforcement, and potential reform of the PIPL provisions. These beg the question of whether the PIPL will be enforced in a way that produces similar outcomes to those of the GDPR. We argue that it will not. Particularly when balanced with other values such as economic growth and security of the party-state, China is likely to depart from the EU model due to the inherent differences in how privacy and data protection are legally formulated and interpreted in practice.

---

[123] Martin Chorzempa and Samm Sacks, 'China's new rules on data flows could signal a shift away from security toward growth' (2023) <https://www.piie.com/blogs/realtime-economics/chinas-new-rules-data-flows-could-signal-shift-away-security-toward-growth> accessed 20 October 2023.

[124] David Gordon and Meia Nowens, 'The Digital Silk Road: Introduction' (2022) <https://www.iiss.org/sv/online-analysis/online-analysis/2022/12/digital-silk-road-introduction/> accessed 20 October 2023.

Overall, the history of China's legal trajectory prior to the GDPR transplantation, when viewed through the lens of Gravity Assist, appears as a deliberate journey characterised by cautious small steps rather than a headlong plunge into comprehensive data protection legislation. This measured approach finds its explanation in several key factors. First, the cultural and historical context of China places a different emphasis on individual rights and values compared to Western societies. Whereas Western societies (particularly Europe after World War II) cherish privacy and data protection as freestanding fundamental rights, China has, for a long history, barely recognised, through legal construction or institution, the significance of privacy. Second, the establishment of an independent, freestanding legal regime for data protection was a daunting task within China's existing legal infrastructure, given the overall status of the rule of law, and its significance at the highest order of values. Third, China's willingness to experiment with areas like consumer protection, cybersecurity, and criminal law demonstrated a pragmatic approach to legal development. The decision to test and adapt various legal instruments without immediately committing to a singular, comprehensive data protection framework was illustrative of its original proposition and consistent with China's own legal culture. As we look into the future, Gravity Assist predicts a trajectory for China's personal information protection law that is significantly shaped, not by the EU regulation and its underlying values, but by its own socio-political agenda. This trajectory may still see some deeper levels of convergence, possibly seen at the principle levels but in the meantime could evolve to the extent that it stands in contrast to the foundational premises of EU data protection law. The cyber sovereignty agenda, the exact meaning of which is constantly evolving, will invariably play a pivotal and predominant role in dictating the direction of data protection in China. It raises fundamental questions about how data privacy will be understood, recognised, and enforced, with the prospect of privacy rights potentially subservient to broader political and economic imperatives in China (e.g., digital innovation, data marketplace, and national security).

While we reflect on China's journey particularly in this paper, the dynamics explored hold implications beyond the PIPL, with the potential to conceptualise similar developments in other non-Western jurisdictions to uncover the future landscape of global data governance. The question of alignment between the PIPL and the GDPR remains pivotal, but the debate should extend beyond mere alignment. The crux of the matter lies in acknowledging the need for a more nuanced lens—a perspective that accounts for both the pull of global influences and the centrifugal forces of internal dynamics. By probing the intricacies and contradictions of China's data protection journey, this paper unveils the need for recalibrating regulatory and theoretical frameworks to accommodate the complex interplay between external influence and local contexts. This recalibration carries profound implications, not only for China's new data protection landscape but also for other non-Western jurisdictions with a similar socio-political climate.

---

[125] Bygrave, '"Strasbourg Effect"' (n 6).

**Acknowledgements**

The authors would like to thank Hunter Stewart, Maria Tzanou, Mark Jia, two anonymous reviewers of the journal, and three additional anonymous reviewers of the International Communications Association (ICA) Annual Conference, for their helpful feedback on a draft of this article. For the purpose of open access, the authors have applied a Creative Commons Attribution (CC BY) licence to any Author Accepted Manuscript version arising.

**References**

Council of Europe Convention for the Protection of Individuals with regard to Automatic Processing of Personal Data ('Council of Europe Convention for the Protection of Individuals with regard to Automatic Processing of Personal Data') [1981] ETS No. 108

Civil Law, General Principles (China) (中华人民共和国民法通则) ('CLGP') [1986]

Directive 95/46/EC of the European Parliament and of the Council of 24 October 1995 on the protection of individuals with regard to the processing of personal data and on the free movement of such data ('Data Protection Directive') [1995] OJ L281/31

Tort Liability Law (China) (中华人民共和国侵权责任法) ('TLT') [2009]

Charter of Fundamental Rights of the European Union ('Charter') [2012] OJ C326/391

Decision of the Standing Committee of the National People's Congress on Preserving Computer Network Security (China) (全国人大常委会关于加强网络信息保护的决定) ('Decision of the Standing Committee of the National People's Congress on Preserving Computer Network Security (China) (全国人大常委会关于加强网络信息保护的决定)') [2012]

National Security Law (China) (中华人民共和国国家安全法) ('NSL') [2015]

Case C-362/14 *Schrems* [2015] OJ C 398/5

Cybersecurity Law (China) (中华人民共和国网络安全法) ('CL') [2016]

Regulation (EU) 2016/679 of the European Parliament and of the Council of 27 April 2016 on the protection of natural persons with regard to the processing of personal data and on the free movement of such data, and repealing Directive 95/46/EC (General Data Protection Regulation) ('GDPR') [2016] OJ L119/1

Civil Code (China) (中华人民共和国民法典) ('CC') [2020]

Case C-311/18 *Facebook Ireland and Schrems* [2020] OJ C 249/21

Case C-319/22 *Gesamtverband Autoteile-Handel (Accès aux informations sur les véhicules)* [2023]

Data Security Law (China) (中华人民共和国数据安全法) ('DSL') [2021]

*Guo v Hangzhou Safari Park (郭兵诉杭州野生动物世界有限公司服务合同纠纷案)* [2021] 浙 01 民终 10940 号

*Huang v Tencent (黄某诉腾讯科技（深圳）有限公司广州分公司、腾讯科技（北京）有限公司隐私权、个人信息权益网络侵权责任纠纷案)* [2019] 京 0491 民初 16142 号

*Ling v TikTok Technology Co., Ltd (凌某某诉北京微播视界科技有限公司隐私权、个人信息权益网络侵权责任纠纷案)* [2019] 京 0491 民初 6694 号

Personal Information Protection Law (China) (中华人民共和国个人信息保护法) ('PIPL') [2021]

Measures for Data Export Security Assessment (China) (数据出境安全评估办法) [2022]

Measures for Promoting and Regulating Data Cross-border Transfers (China) (促进和规范数据跨境流动规定) [2024]

*Yu v Chaboshi (余某某诉查博士)* [2021] 粤 0192 民初 928 号

*Zhou v Vipshop (周彦聪诉广州唯品会电子商务有限公司个人信息保护纠纷案)* [2022] 粤 01 民终 3937 号

Adams J and Almahmoud H, 'The Meaning of Privacy in the Digital Era' (2023) 15(1) International Journal of Security and Privacy in Pervasive Computing (IJSPPC) 1

Albrecht D, 'New Developments in Chinese Data Protection Law in Contrast to the European GDPR—New requirements for network operators in China' (2021) 22(5) Computer Law Review International 142

Baasanjav UB, Fernback J and Pan X, 'A critical discourse analysis of the human flesh search engine' (2019) 46(1-2) Media Asia 18

Bennett CJ, *Regulating privacy: Data protection and public policy in Europe and the United States* (Cornell University Press 1992)

Bennett CJ and Raab CD, *The Governance of Privacy: Policy Instruments in Global Perspective* (The MIT Press 2006)

Bradford A, 'The Brussels Effect' (2012) 107(1) Northwestern University Law Review 1

——, 'Exporting standards: The externalization of the EU's regulatory power via markets' (2015) 42 International Review of Law and Economics 158

——, *The Brussels Effect: How the European Union Rules the World* (Oxford University Press 2020)

——, *Digital Empires: The Global Battle to Regulate Technology* (Oxford University Press 2023)

Broucke RA, 'The celestial mechanics of gravity assist' (Astrodynamics Conference, Minneapolis, August 1988)

Bygrave LA, 'Privacy and data protection in an international perspective' (2010) 56(8) Scandinavian studies in law 165

——, *Data Privacy Law: An International Perspective* (Oxford University Press 2014)

——, 'The "Strasbourg Effect" on data protection in light of the "Brussels Effect": Logic, mechanics and prospects' (2021) 40 Computer Law & Security Review 105460

Chander A and Schwartz PM, 'Privacy and/or Trade' (2023) 90(1) University of Chicago Law Review 49

Chen J and others, 'Who Is Responsible for Data Processing in Smart Homes? Reconsidering Joint Controllership and the Household Exemption' (2020) 10(4) International Data Privacy Law 279

Chorzempa M and Sacks S, 'China's new rules on data flows could signal a shift away from security toward growth' (2023) <https://www.piie.com/blogs/realtime-economics/chinas-new-rules-data-flows-could-signal-shift-away-security-toward-growth> accessed 20 October 2023

Clarke D, 'Anti anti-Orientalism, or is Chinese law different?' (2020) 68(1) The American Journal of Comparative Law 55

Commission, 'EU-China Comprehensive Agreement on Investment' (2020) <https://policy.trade.ec.europa.eu/eu-trade-relationships-country-and-region/countries-and-regions/china/eu-china-agreement_en> accessed 20 October 2023

——, 'Commission opens formal proceedings against TikTok under the Digital Services Act' (2024) <https://ec.europa.eu/commission/presscorner/detail/en/ip_24_926> accessed 5 May 2024

Creemers R, 'China's conception of cyber sovereignty' in Broeders D and van den Berg B (eds), *Governing cyberspace: Behavior, power and diplomacy* (2020)

——, 'China's emerging data protection framework' (2022) 8(1) Journal of Cybersecurity 1

Cui H and Fan C, 'A Study On Judicial Application Of Crime Of Infringing Citizens' Personal Information: Based On Some Judicial Decisions In Beijing (侵犯公民个人信息罪的司法适用研究——以北京市部分司法判决为基础)' (2022)(5) Journal of Shandong Police College 26

Data Protection Commission, 'Irish Data Protection Commission announces €345 million fine of TikTok' (2023) <https://www.dataprotection.ie/en/news-media/press-releases/DPC-announces-345-million-euro-fine-of-TikTok> accessed 20 October 2023

De Hert P and Gutwirth S, 'Data protection in the case law of Strasbourg and Luxemburg: Constitutionalisation in action' in Gutwirth S and others (eds), *Reinventing data protection?* (Springer 2009)

De Hert P and Papakonstantinou V, *The data protection regime in China: In-depth analysis for the LIBE Committee* (2015) <https://www.europarl.europa.eu/RegData/etudes/IDAN/2015/536472/IPOL_IDA(2015)536472_EN.pdf> accessed 20 October 2023

Deng J and Dai K, 'The Comparison Between China's PIPL And EU's GDPR: Practitioners' Perspective' (2021) <https://www.mondaq.com/china/data-protection/1122748/the-comparison-between-chinas-pipl-and-eus-gdpr-practitioners-perspective> accessed 20 October 2023

Ding X, 'Reflection and Reconstruction of Personal Information Rights: On the Applicable Premise and Legal Interest Basis of Personal Information Protection Law (个人信息权利的反思与重塑——论个人信息保护的适用前提与法益基础)' (2020) 32(2) Peking University Law Journal 339

Eurostat, 'International trade in services (since 2010) (BPM6)' (2023) <https://ec.europa.eu/eurostat/databrowser/bookmark/7103afb1-9e06-4017-8939-2bcb9e5a9d10?lang=en> accessed 20 October 2023

Gao F, 'On the Purpose of Personal Information Protection: A Perspective of the Protected Interests on Personal Information (论个人信息保护的目的——以个人信息保护法益区分为核心)' (2019) 36(1) Civil and Commercial Law 93

Gao Y and Xu J, 'Unclear Supervisors Behind App Personal Information Protection' (2022) <https://www.dehenglaw.com/CN/tansuocontent/0008/025360/7.aspx?MID=0902> accessed 20 October 2023

Geller A, 'How Comprehensive Is Chinese Data Protection Law? A Systematisation of Chinese Data Protection Law from a European Perspective' (2020) 69(12) GRUR International 1191

Ginsburg T, 'Does Law Matter for Economic Development? Evidence from East Asia' (2000) 34 Law and Society Review 829

Gordon D and Nowens M, 'The Digital Silk Road: Introduction' (2022) <https://www.iiss.org/sv/online-analysis/online-analysis/2022/12/digital-silk-road-introduction/> accessed 20 October 2023

Graef I, Clifford D and Valcke P, 'Fairness and enforcement: bridging competition, data protection, and consumer law' (2018) 8(3) International Data Privacy Law 200

Greenleaf G, 'The influence of European data privacy standards outside Europe: implications for globalization of Convention 108' (2012) 2(2) International Data Privacy Law 68

——, *Asian Data Privacy Laws: Trade and Human Rights Perspectives* (Oxford University Press 2014)

——, 'The 'Brussels effect'of the EU's 'AI Act'on data privacy outside Europe' <https://ssrn.com/abstract=3898904> accessed 20 October 2023

——, 'China's Completed Personal Information Protection Law: Rights Plus Cyber-security' <https://dx.doi.org/10.2139/ssrn.3989775> accessed 20 October 2023

——, 'Global Data Privacy 2023: DPA Networks Almost Everywhere' <https://dx.doi.org/10.2139/ssrn.4461729> accessed 20 October 2023

——, 'Global data privacy laws 2023: 162 national laws and 20 Bills' <https://dx.doi.org/10.2139/ssrn.4426146> accessed 20 October 2023

——, 'Global data privacy laws 2023: International standards stall, but UK disrupts' <https://dx.doi.org/10.2139/ssrn.4530145> accessed 20 October 2023

——, 'Global data privacy laws: EU leads US and the rest of the world in enforcement by penalties' <https://dx.doi.org/10.2139/ssrn.4409491> accessed 20 October 2023

Greenleaf G and Livingston S, 'China's New Cybersecurity Law–Also a Data Privacy Law?' <https://ssrn.com/abstract=2958658> accessed 20 October 2023

——, 'China's personal information standard: the long march to a privacy law' <https://ssrn.com/abstract=3128593> accessed 20 October 2023

Han D, 'Search boundaries: human flesh search, privacy law, and internet regulation in China' (2018) 28(4) Asian Journal of Communication 434

Helberger N, Borgesius FZ and Reyna A, 'The perfect match? A closer look at the relationship between EU consumer law and data protection law' (2017) 54(5) Common Market Law Review 1427

Jayasuriya K, 'The rule of law and governance in East Asia' in Mark Beeson (ed) Reconfiguring East Asia (Routledge 2002) 99-116

Jia M, 'Authoritarian Privacy' <https://dx.doi.org/10.2139/ssrn.4362527> accessed 20 October 2023

Ke X and others, 'Analyzing China's PIPL and how it compares to the EU's GDPR' <https://iapp.org/news/a/analyzing-chinas-pipl-and-how-it-compares-to-the-eus-gdpr> accessed 20 October 2023

Kennedy G and Woo JTK, 'The CAC is Coming: Didi Chuxing Fined a Record-breaking USD 1.2 Billion for Breach of Data Protection Regulations' (2022) <https://www.mayerbrown.com/en/perspectives-events/publications/2022/08/the-cac-is-comingdidi-chuxing-fined-a-recordbreaking-usd-12-billion-for-breach-of-data-protection-regulations> accessed 20 October 2023

Kuner C, 'The Internet and the Global Reach of EU Law' in Cremona M and Scott J (eds), *EU Law Beyond EU Borders: The Extraterritorial Reach of EU Law* (Oxford University Press 2019)

Laird J, 'The GDPR vs China's PIPL' (2022) <https://www.privacypolicies.com/blog/gdpr-vs-pipl/> accessed 20 October 2023

Li T, Bronfman J and Zhou Z, 'Saving Face: Unfolding the Screen of Chinese Privacy Law' <https://ssrn.com/abstract=2826087> accessed 20 October 2023

Liu Q, 'Disagreements on the Rights of Personal Information Protection and Resolution (个人信息保护的权利化分歧及其化解)' (2022) 52(6) China Law Review 107

Lü Y-H, 'Privacy and data privacy issues in contemporary China' (2005) 7 Ethics and Information Technology 7

Lynskey O, *The Foundations of EU Data Protection Law* (Oxford University Press 2015)

——, 'Article 84 Penalties' in Kuner C and others (eds), *The EU General Data Protection Regulation (GDPR): A Commentary* (Oxford University Press 2020)

Mannion C, 'Data imperialism: The GDPR's disastrous impact on Africa's E-commerce markets' (2020) 53 Vanderbilt Journal of Transnational Law 685

MIIT, 'Notice on APPs (SDK) Infringing User Interest (关于侵害用户权益行为的 APP（SDK）通报（2024 年第 2 批，总第 37 批）)' (2024) <https://wap.miit.gov.cn/xwdt/gxdt/sjdt/art/2024/art_424e1faa0479457689f7a49e113a5429.html> accessed 5 May 2024

Noerr, 'PIPL: Data protection made in China' (2021) <https://www.noerr.com/en/insights/pipl-data-protection-made-in-china> accessed 20 October 2023

OECD, 'OECD Guidelines on the Protection of Privacy and Transborder Flows of Personal Data' (1980) <https://read.oecd-ilibrary.org/science-and-technology/oecd-guidelines-on-the-protection-of-privacy-and-transborder-flows-of-personal-data_9789264196391-en> accessed 13 June 2020

Ong R, 'Recognition of the right to privacy on the Internet in China' (2011) 1(3) International Data Privacy Law 172

Peerenboom R, 'Globalization, path dependency and the limits of law: Administrative law reform and rule of law in the People's Republic of China' (2001) 19 Berkeley Journal of International Law 161

Pernot-LePlay E, 'China's approach on data privacy law: A third way between the US and the EU?' (2020) 8(1) Penn State Journal of Law & International Affairs 49

Polonetsky J, Tene O and Finch K, 'Shades of gray: Seeing the full spectrum of practical data de-intentification' (2016) 56 Santa Clara Law Review 593

Potter A, Campbell K and Ashcroft V, 'Comparing Privacy Laws: GDPR v. PIPL' (2022) <https://www.dataguidance.com/resource/comparing-privacy-laws-gdpr-v-pipl> accessed 20 October 2023

Qi A, 'Scholarly Proposal of the Personal Information Protection Law (中华人民共和国个人信息保护法学者建议稿)' (2019) 37(1) Hebei Law Science 33

Qi A and Zhang Z, 'Identification and reidentification: The definition of personal information and the legislative choice (识别与再识别: 个人信息的概念界定与立法选择)' (2018) 24(2) Journal of Chongqing University (Social Science Edition) 119

Reuters, 'China looks to increase penalties under its cybersecurity law' (2022) <https://www.reuters.com/world/china/china-seeks-public-comment-possible-amendments-cybersecurity-law-2022-09-14/> accessed 20 October 2023

Segal A, 'When China rules the web: technology in service of the state' (2018) 97 Foreign Affairs 10

Shen W, 'On the Construction and Systematization of the Personal Information Right (论个人信息权的构建及其体系化)' (2021) 189(5) Journal of Comparative Law 1

Sheng X and Tang J, 'A Comparative Analysis of Personal Information Rights in China and Personal Data Rights in EU: Based on the Personal Information Protection Law and GDPR (我国个人信息权利与欧盟个人数据权利的比较分析：基于《个人信息保护法》与 GDPR)' (2022) 66(6) Library and Information Service 26

Siegmann C and Anderljung M, 'The Brussels effect and artificial intelligence: How EU regulation will impact the global AI market' <https://doi.org/10.48550/arXiv.2208.12645> accessed 20 October 2023

Sohu, 'Personal Information Protection Law of People's Republic of China (Draft) 2017 (中华人民共和国个人信息保护法（草案）2017 版)' (2017) <https://www.sohu.com/a/203902011_500652> accessed 20 October 2023

Solove D, 'China's PIPL vs. the GDPR: A Comparison' (2021) <https://teachprivacy.com/chinas-pipl-vs-gdpr-a-comparison> accessed 23 October 2023

Stanford Law School, 'China Guiding Cases Project' (2021) <https://law.stanford.edu/china-guiding-cases-project> accessed 20 October 2023

The People's Bank of China, 'Notice of the People's Bank of China on the Public Consultation on Banking Sector Data Security Management Measures (Consultation Draft) (中国人民银行关于《中国人民银行业务领域数据安全管理办法（征求意见稿）》公开征求意见的通知)' (2023) <www.pbc.gov.cn/tiaofasi/144941/144979/3941920/4993510/index.html> accessed 20 October 2023

Vogel D, *Trading up: Consumer and environmental regulation in a global economy* (Harvard University Press 1995)

Waldersee V, 'Volkswagen China chief asks China's premier Li for clarity on data transfers' (2023) <https://www.reuters.com/business/autos-transportation/volkswagen-china-chief-asks-chinas-premier-li-clarity-data-transfers-2023-06-27/> accessed 20 October 2023

Wang A, 'Cyber Sovereignty at Its Boldest: A Chinese Perspective' (2020) 16 Ohio State Technology Law Journal 395

Wang S, 'Authoritarian Legality and Legal Instrumentalism in China' (2022) 10(1) The Chinese Journal of Comparative Law 154

Wang X, 'The Three-level Frame and Protection Mechanism of Personal Information Rights and Interests (个人信息权益的三层构造及保护机制)' (2021) 43(5) Modern Law Science 105

Wu Q, 'How has China formed its conception of the rule of law? A contextual analysis of legal instrumentalism in ROC and PRC law-making' (2017) 13(3) International Journal of Law in Context 277

Xue H, 'Privacy and personal data protection in China: An update for the year end 2009' (2010) 26(3) Computer Law & Security Review 284

You C, 'Half a loaf is better than none: The new data protection regime for China's platform economy' (2022) 45 Computer Law & Security Review 105668

Zhang L and Xu D, 'On the legal regulation of personal information anonymizationin China under the risk control concept (风险控制理念下我国个人信息匿名化处理的法律规制)' (2023) 29(2) Journal of Chongqing University (Social Science Edition) 220

Zhang T, 'GDPR Versus PIPL – Key Differences and Implications for Compliance in China' (2022) <https://www.china-briefing.com/news/pipl-vs-gdpr-key-differences-and-implications-for-compliance-in-china> accessed 20 October 2023

Zhang X, 'A Discussion on the Legislation of Individual Information Protection of China (我国个人信息保护法立法主要矛盾研讨)' (2018) 58(5) Jilin University Journal Social Sciences Edition 45

——, 'From Privacy to Personal Information: Theory and Institutional Arrangement of Interest Rebalance (从隐私到个人信息: 利益再衡量的理论与制度安排)' [2015] Peking University Law Journal 38

——, 'The Structure of Rights and Interests Relating to Personal Information (论个人信息权益的构造)' (2021) 33(5) Peking University Law Journal 1144

Zhou H, *Personal Information Protection Law (Expert Proposal) and Legislative Research Report (个人信息保护法（专家建议稿）及立法研究报告)* (Law Press China 2006)

Zhu C, 'Is China's New Personal Information Privacy Law the New GDPR?' (2021) <https://news.bloomberglaw.com/privacy-and-data-security/is-chinas-new-personal-information-privacy-law-the-new-gdpr> accessed 20 October 2023

Zhu G, 'The Right to Privacy: An Emerging Right in Chinese Law' (1997) 18(3) Statute Law Review 208

Zhu J, 'The Personal Information Protection Law: China's Version of the GDPR' <https://www.jtl.columbia.edu/bulletin-blog/the-personal-information-protection-law-chinas-version-of-the-gdpr> accessed 20 October 2023